\documentclass[11pt]{article}
\usepackage{graphicx}

\newcommand{\BABARPubYear}    {00}
\newcommand{\BABARPubNumber}  {17}
\newcommand{\SLACPubNumber} {8539}

\usepackage{epsfig}

\usepackage{relsize}
\def\babar{\mbox{\slshape B\kern-0.1em{\smaller A}\kern-0.1em
    B\kern-0.1em{\smaller A\kern-0.2em R}}}

\def\electron   {\ensuremath{e}}
\def\en         {\ensuremath{e^-}}      

\def\epm        {\ensuremath{e^\pm}}  
\def\epem       {\ensuremath{e^+e^-}}

\def\mumu       {\ensuremath{\mu^+\mu^-}}

\def\gaga  {\ensuremath{\gamma\gamma}}  

\def\ccbar {\ensuremath{c\overline c}}

\def\bbbar {\ensuremath{b\overline b}}

\def\piz   {\ensuremath{\pi^0}}

\def\ppz   {\ensuremath{\pi^0\pi^0}}
\def\pip   {\ensuremath{\pi^+}}
\def\pim   {\ensuremath{\pi^-}}
\def\pipi  {\ensuremath{\pi^+\pi^-}}

\def\Kbar  {\kern 0.2em\overline{\kern -0.2em K}{}}

\def\Kp    {\ensuremath{K^+}}
\def\Km    {\ensuremath{K^-}}
\def\KS    {\ensuremath{K^0_{\scriptscriptstyle S}}} 
\def\KL    {\ensuremath{K^0_{\scriptscriptstyle L}}}

\def\Kzb   {\ensuremath{\Kbar^0}}
\def\KzKzb {\ensuremath{K^0 \kern -0.16em \Kzb}}
\def\Dz    {\ensuremath{D^0}}
\def\Dbar  {\kern 0.2em\overline{\kern -0.2em D}{}}

\def\Dzb   {\ensuremath{\Dbar^0}}
\def\DzDzb {\ensuremath{D^0 {\kern -0.16em \Dzb}}}

\def\Bz    {\ensuremath{B^0}}
\def\B     {\ensuremath{B}}
\def\Bbar  {\kern 0.18em\overline{\kern -0.18em B}{}}

\def\Bzb   {\ensuremath{\Bbar^0}}

\def\Bub   {\ensuremath{B^-}}

\def\BB    {\ensuremath{B\Bbar}} 
\def\BzBzb {\ensuremath{B^0 {\kern -0.16em \Bzb}}}

\def\jpsi  {\ensuremath{{J\mskip -3mu/\mskip -2mu\psi\mskip 2mu}}} 
\mathchardef\Upsilon="7107
\def\Y#1S{\ensuremath{\Upsilon{(#1S)}}}

\def\FourS {\Y4S}

\mathchardef\Deltares="7101
\mathchardef\Xi="7104
\mathchardef\Lambda="7103
\mathchardef\Sigma="7106
\mathchardef\Omega="710A
\def\Deltabar   {\kern 0.25em\overline{\kern -0.25em \Deltares}{}}
\def\Lbar {\kern 0.2em\overline{\kern -0.2em\Lambda\kern 0.05em}\kern-0.05em{}}
\def\Sigbar{\kern 0.2em\overline{\kern -0.2em \Sigma}{}}
\def\Xibar{\kern 0.2em\overline{\kern -0.2em \Xi}{}}
\def\Obar{\kern 0.2em\overline{\kern -0.2em \Omega}{}}
\def\Nbar{\kern 0.2em\overline{\kern -0.2em N}{}}
\def\Xbar{\kern 0.2em\overline{\kern -0.2em X}{}}

\def\bpsiks{\ensuremath{B^0 \to \jpsi \KS}}

\def\pt         {\mbox{$p_T$}}
\def\mes        {\mbox{$m_{\rm ES}$}}
\def\mec        {\mbox{$m_{\rm EC}$}}

\def\ev   {\ensuremath{\rm \,e\kern -0.08em V}}
\def\kev  {\ensuremath{\rm \,ke\kern -0.08em V}} 
\def\mev  {\ensuremath{\rm \,Me\kern -0.08em V}} 
\def\gev  {\ensuremath{\rm \,Ge\kern -0.08em V}} 
\def\gevc {\ensuremath{{\rm \,Ge\kern -0.08em V\!/}c}} 
\def\tev  {\ensuremath{\rm \,Te\kern -0.08em V}}
\def\mevc {\ensuremath{{\rm \,Me\kern -0.08em V\!/}c}} 
\def\gevcc{\ensuremath{{\rm \,Ge\kern -0.08em V\!/}c^2}} 
\def\mevcc{\ensuremath{{\rm \,Me\kern -0.08em V\!/}c^2}}

\def\km   {\ensuremath{\rm \,km}}
\def\m    {\ensuremath{\rm \,m}}
\def\cm   {\ensuremath{\rm \,cm}}
\def\cma  {\ensuremath{{\rm \,cm}^2}}
\def\mm   {\ensuremath{\rm \,mm}}

\def\mum  {\ensuremath{\,\mu\rm m}} 

\def\pb {\ensuremath{\rm \,pb}}

\def\invfb   {\ensuremath{\mbox{\,fb}^{-1}}}
\def\mus  {\ensuremath{\rm \,\mus}}
\def\ns   {\ensuremath{\rm \,ns}}

\def\mus        {\ensuremath{\,\mu{\rm s}}}    
\def\ns         {\ensuremath{{\rm \,ns}}}      
\def\degc {\ensuremath{^\circ}{C}}

\def\degrees{\ensuremath{^{\circ}}}
\def\mrad{\ensuremath{\rm \,mr}}                

\def\gsim{{~\raise.15em\hbox{$>$}\kern-.85em
          \lower.35em\hbox{$\sim$}~}}
\def\lsim{{~\raise.15em\hbox{$<$}\kern-.85em
          \lower.35em\hbox{$\sim$}~}}

\def\CP                 {\ensuremath{C\!P}}
\def\ra                 {\ensuremath{\rightarrow}}
\def\to                 {\ensuremath{\rightarrow}}

\def\pep2{PEP-II}
\def\BF{$B$ Factory}

\def\chic#1{\ensuremath{\chi_{c#1}}} 

\newcommand{\dedx}{\ensuremath{\mathrm{d}\hspace{-0.1em}E/\mathrm{d}x}}

\def\stwob{\ensuremath{\sin\! 2 \beta   }}

\providecommand{\eqref}[1]{Eq.~(\ref{eq:#1})}

\newcommand{\nim}       [1]  {{Nucl.\ Instr.\ and Methods~{\bf #1}}}

\providecommand{\pl}        [1]  {{Phys.\ Lett.\ {\bf #1}}}      

\providecommand{\pr}        [1]  {{Phys.\ Rev.\ {\bf #1}}}

\newcommand{\zp}        [1]  {{Z.\ Phys.\ {\bf #1}}}

\def\bbsim      {\mbox{\tt BBsim}}

\def\evtgen     {\mbox{\tt EvtGen}}

\def\fluka          {\mbox{\tt Fluka}}

\def\gcalor     {\mbox{\tt GCalor}}
\def\geant      {\mbox{\tt Geant321}}
\def\gheisha    {\mbox{\tt Gheisha}}
\def\jetset74   {\mbox{\tt Jetset \hspace{-0.5em}7.\hspace{-0.2em}4}}

\def\bbsim      {\mbox{\tt BBsim}}
\def\SimApp     {\mbox{\tt SimApp}}

\def\evtgen     {\mbox{\tt EvtGen}}
\def\fluka	{\mbox{\tt Fluka}}

\def\gcalor     {\mbox{\tt GCalor}}
\def\geant      {\mbox{\tt Geant321}}
\def\gheisha    {\mbox{\tt Gheisha}}
\def\jetset74   {\mbox{\tt Jetset \hspace{-0.5em}7.\hspace{-0.2em}4}}
\def\jetset     {\mbox{\tt Jetset \hspace{-0.5em}7.\hspace{-0.2em}4}}

\def\bzb   {\ensuremath{\Bbar^0}}

\def\BB    {\ensuremath{B\Bbar}} 
\def\BzBzb {\ensuremath{B^0 \Bbar^0}} 
\def\Bbar  {{\kern 0.18em\overline{\kern -0.18em B}}{}}

\long\def\inst#1{\par\nobreak\kern 4pt\nobreak
    {\it #1}\par\vskip 10pt plus 3pt minus 3pt}

\begin{document}
{\pagestyle{empty}

\begin{flushright}
\babar-CONF-\BABARPubYear/\BABARPubNumber \\
SLAC-PUB-\SLACPubNumber \\
\end{flushright}

\par\vskip 3cm

\begin{center}
\Large \bf
The first year of  
the \babar\ experiment at \pep2
\end{center}
\bigskip

\begin{center}
\large The \babar\ Collaboration\\
\mbox{ }\\
\today
\end{center}
\bigskip 



\begin{center}
\large \bf Abstract
\end{center}
The \babar\ detector, situated at the SLAC \pep2\ asymmetric \epem\ 
collider, has been recording data at energies on and around the \FourS\ resonance since
May 1999.  In this paper, we briefly describe the \pep2\ \BF\ and the 
\babar\ detector.  The performance presently achieved by the experiment
in the areas of tracking, vertexing, 
calorimetry and particle identification
is reviewed.
Analysis concepts that are used in the various
papers submitted to this conference are also discussed.

\vfill
\centerline
{Submitted to the XXX$^{th}$ International Conference on High Energy Physics, Osaka, Japan}
\newpage
}

\newcommand{\secname}{}

\begin{center}
\small

The \babar\ Collaboration
\bigskip

B.~Aubert,
A.~Boucham,
D.~Boutigny,
I.~De Bonis,
J.~Favier,
J.-M.~Gaillard,
F.~Galeazzi,
A.~Jeremie,
Y.~Karyotakis,
J.~P.~Lees,
P.~Robbe,
V.~Tisserand,
K.~Zachariadou
\inst{Lab de Phys.\ des Particules, F-74941 Annecy-le-Vieux, CEDEX, France}
% ...

\end{center}\newpage

\newpage
\tableofcontents
\newpage

\renewcommand{\secname}{Intro}          
\section{Introduction} 
\label{sec:\secname}


The principal goal of an asymmetric \BF\ running at energies on and around the \FourS\ resonance
is the comprehensive study of \CP\ violation in \B\ meson systems.
The asymmetry in beam energies makes possible the measurement of time-dependent
\CP\ violating asymmetries in the decay of neutral \B\ mesons.
\par
\CP\ asymmetries in the \B\ meson system are expected to be large, so that
relatively small samples of events are needed to provide accurate measurements.
Unfortunately, the \B\ meson decay channels of interest for the study of \CP\
violation have extremely small branching fractions, of order of $10^{-4}$ or below.
In order to measure asymmetries with errors at the 10\% level or better, samples
of several tens of million neutral \B\ meson pairs are needed. The new
generation of \B\ factories must therefore have unprecedented luminosities, in the range
$10^{33}$ to $10^{34}\cm^{-2}s^{-1}$.

\renewcommand{\secname}{PEPII}          
\section{The \pep2\ \BF}
\label{sec:\secname}



The \pep2\ \BF\ is an \epem\ colliding beam storage ring complex designed to produce
a luminosity of at least $3\times 10^{33} \cm^{-2}s^{-1}$ at $\rm 
{E_{CM}=10.58}$\gev, the
mass of the \FourS\ resonance. \pep2\ has been constructed by a collaboration 
of
SLAC, LBNL and LLNL~\cite{PEP_II} 
on the SLAC site. The machine is asymmetric with
energies of 9.0\gev\ for the electron  beam and 3.1\gev\ for the positron 
beam.
The unequal beam energies require a two-ring configuration:  electrons in
a High Energy Ring (HER) colliding with  positrons in the  Low
Energy Ring (LER). Some accelerator parameters and achievements
of the High Energy Ring are listed in Table~\ref{tab:HER_perf}, those of
the Low Energy Ring in Table~\ref{tab:LER_perf}. The electrons and positrons
are produced by the SLAC Linac, whose high intensity makes it optimal  to
recharge the beams in ``top-off'' mode  whenever the luminosity  drops   to
about 90\% of its peak value. 

\begin{table}[!htb]
\caption{\pep2\ High Energy Ring Performance Results.}
\begin{center}
\begin{tabular}{@{}lcccc@{}}\hline
Parameter          & units   & Design     &Best Achieved   &Typical in-run   
\\
\hline
Energy             & GeV     &  9.0       & 9.0, ramp to 9.1 $\&$ back& 9.0,
ramp 8.84-9.04 \\
Number of Bunches  &         & 1658       & 1658           & 553-829         
\\
Total Beam Current & A       & 0.75 (1.0) & 0.92           & 0.70            
\\
Beam Lifetime      &         & 4hrs @ 1A  & 11hrs @ 0.9A   & 9hrs @ 0.70A    
\\	   		     
\hline
\end{tabular}
\end{center}
\label{tab:HER_perf}
\end{table}

\begin{table}[!htb]
\caption{\pep2\ Low Energy Ring Performance Results.}
\begin{center}
\begin{tabular}{@{}lcccc@{}}\hline
Parameter          & units   & Design     &Best Achieved   &Typical in-run   
\\
\hline
Energy             & GeV     &  3.1       & 3.1            & 3.1             
\\
Number of Bunches  &         & 1658       & 1658           & 553-829         
\\
Total Beam Current & A       & 2.14       & 1.72           & 1.10            
\\
Beam Lifetime      &         & 4hrs @ 2A  & 3.3hrs @ 1.4A  & 3hrs @ 1.1A     
\\	   		     
\hline
\end{tabular}
\end{center}
\label{tab:LER_perf}
\end{table}
 
The rings are housed in the 2.2\km\ former
PEP tunnel but with distinct vacuum and accelerating structures. The
High Energy Ring (HER) reuses
the magnets of the old PEP machine whereas the Low Energy Ring (LER) is new 
and is put in place
on top of the HER.  The \pep2\ design has 1658 bunches,
each containing $2.1\times 10^{10}$ electrons (HER) and
$5.9\times 10^{10}$ positrons (LER),
spaced at 4.2\ns. The RF system provides a total power of 5.1~MW from seven 
klystron
stations driving 24 conventional copper 476~MHz RF cavities. Bunches are 
brought
into a common vacuum chamber and septum at a few nTorr and into head-on
collisions in an  interaction region  within a  2.5\cm-radius beryllium
beam-pipe, around  which \babar\
is located. 

Asymmetric collisions produce a center of mass energy of 10.58\gev\ with a 
boost of
$\beta\gamma=0.56$ in the lab frame, thereby allowing a
measurement of time dependent \CP\ violating asymmetries. Such a boost 
produces an
average separation of $\rm {\beta\gamma c \tau} =250$\mum\ between the two \B\
vertices, which is crucial for studying the cleanest and most promising \CP\
violating modes.
 
The \pep2\ \epem\ collider became operational in July 1998 with the completion 
of the LER.
The first collisions were seen shortly thereafter. Fall and winter 1998
\pep2\ runs concentrated on raising the beam currents and increasing the
luminosity. In February 1999, the peak luminosity reached $5.2\times 10^{32}
\cm^{-2}s^{-1}$. In a two month spring down time, the \babar\ detector was
installed. \pep2\ turned on May $10^{th}$ and \babar\ saw its first hadronic 
event
on May $26^{th}$ 1999. In August 1999, \pep2\ passed the world record for
luminosity, achieving $8.1\times 10^{32}\cm^{-2}s^{-1}$. The present 
luminosity in
\pep2\ is $2.17\times 10^{33}\cm^{-2}s^{-1}$, or about 72\% of the design 
goal. In June
2000, \pep2\ delivered an integrated luminosity of 150\pb$^{-1}$ in one day,
above the design goal for daily integrated luminosity of 135\pb$^{-1}$. 
Through July 6, \pep2\ has delivered over 13.8\invfb\ to \babar, of which 
12.7\invfb\ have
been logged by the detector. Beam parameters and luminosity
performance results are shown in Table~\ref{tab:lumi_perf}. The present plan 
is 
to collide
until the end of October 2000 followed by a three month installation period. A
detailed status report of the current \pep2\ performance can be found in 
Ref.~\cite{Seeman}.
 
\begin{table}[!htb]
\caption{\pep2\ Luminosity Performance.}
\begin{center}
\begin{tabular}{@{}llccc@{}}\hline
Parameter        &   units         & Design       &Best Achieved &Typical in-
run\\
\hline
Peak Luminosity  & $\cm^{-2}s^{-1}$& $3\times 10^{33}$   & $2.2\times 10^{33}$ 
& $2.0\times 10^{33}$\\
Specific Luminosity& $\cm^{-2}s^{-1}$/bunch& $3.1\times 10^{30}$ & $2.9\times 
10^{30}$ & $1.8\times 10^{30}$\\
                 &                 &              &             &             
\\
IP Hor. Spot Size $\Sigma_X$& \mum\           &   220        &   190       &             
\\
IP Ver. Spot Size $\Sigma_Y$& \mum\           &    6.7       &    6.0      &             
\\
                 &                 &              &             &             
\\
Top-off Time     &   minutes       &     3        &     2       &      3      
\\
Fill Time        &   minutes       &     6        &     8       &     10      
\\
                 &                 &              &             &             
\\
Integrated Luminosity & pb$^{-1}$/8 hours&            &    54     &             
\\   
                 & pb$^{-1}$/day   &     135      &     154     &             
\\
                 & fb$^{-1}$/week  &              &      0.8    &             
\\
		 & fb$^{-1}$/month &              &      2.7    &             \\
Total Integrated Luminosity  &   fb$^{-1}$ &       &  15.4   &      \\
(thru 7/20/2000)	&     &     &     &      \\	   		     
\hline
\end{tabular}
\end{center}
\label{tab:lumi_perf}
\end{table}

\begin{figure}[htb]
\begin{center}
\includegraphics[width=2.9in]{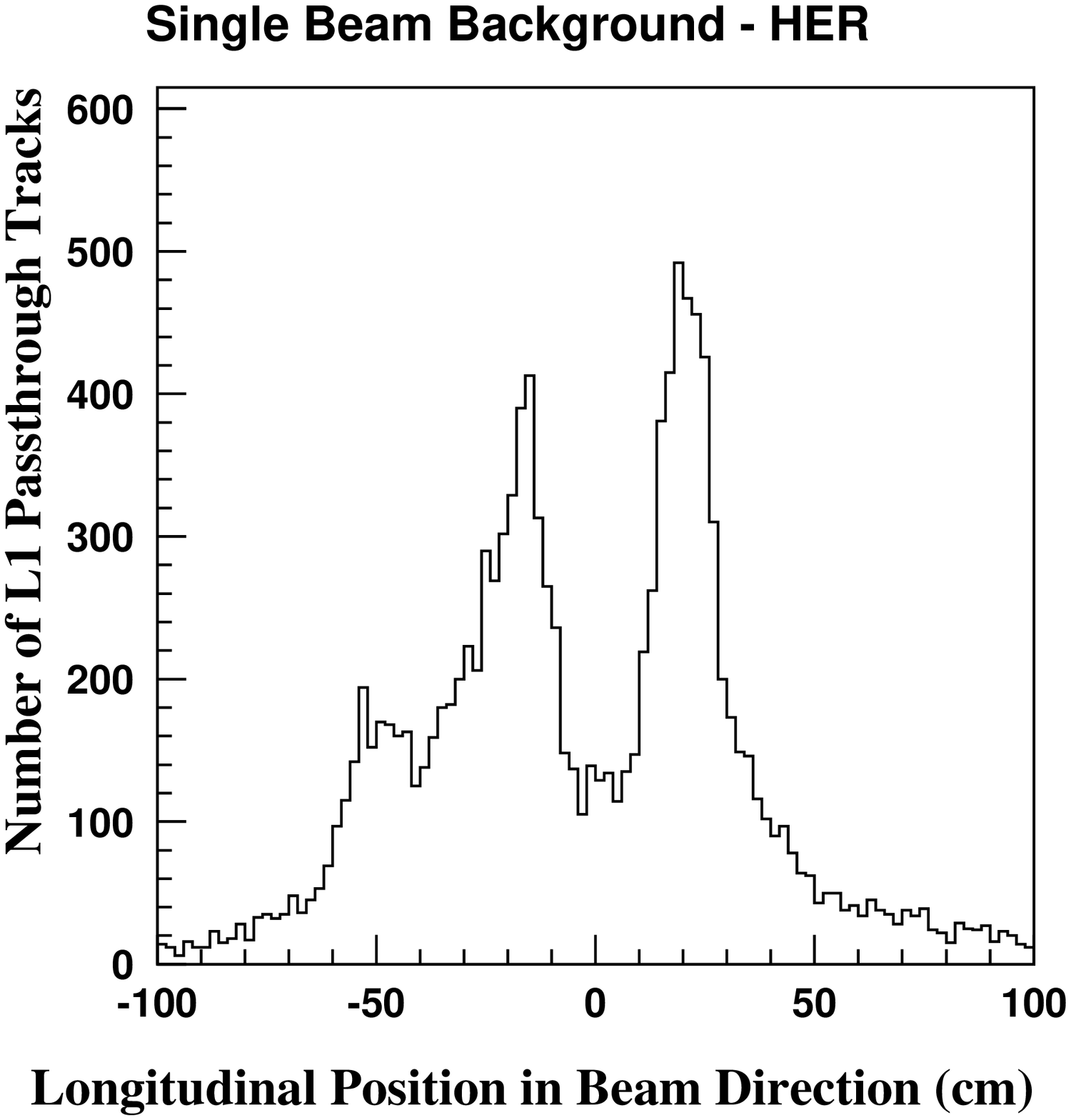}
\includegraphics[width=2.9in]{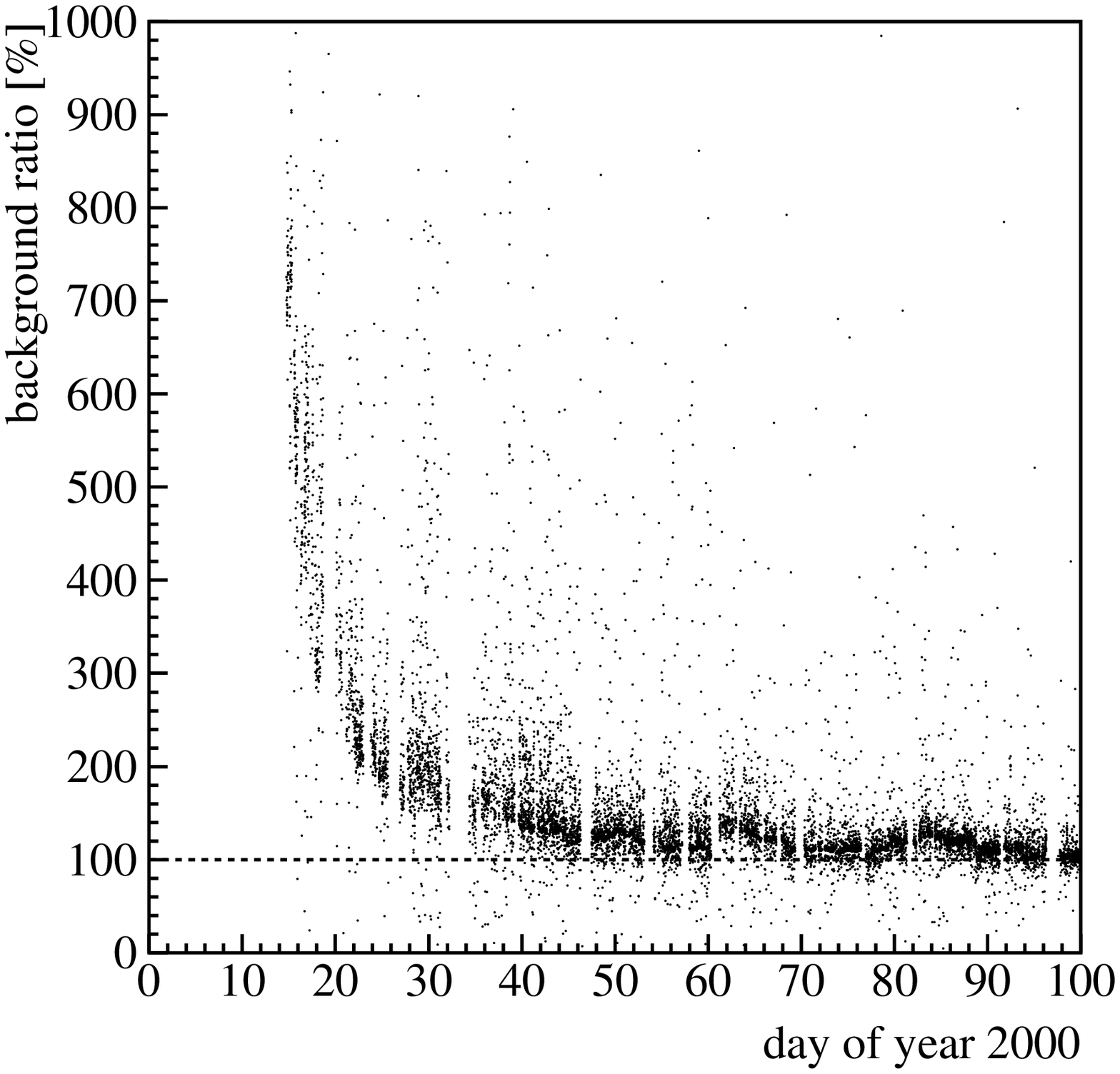}
\end{center}
\caption{Left: $z_{0}$ distribution of L1 pass-through tracks at their point 
of closest approach to the beam line.
Right: Time-history of the DCH current, normalized to that expected, 
at the same beam currents and luminosity, from background measurements 
performed in July 2000. The exponential decrease during the first two months 
of operation reflects the scrubbing, by synchrotron-radiation photons, of 
residual gas molecules off the vacuum-pipe wall.
} 
\label{fig:L1vtx}
\end{figure}
 
\subsection{Machine Backgrounds}
\label{sec:\secname:Backgrounds}

Operationally, the acceptable level of backgrounds is determined
primarily by the radiation hardness of the Silicon Vertex Tracker (SVT) and
Electromagnetic Calorimeter (EMC) detectors, and by requiring a
tolerable Drift Chamber (DCH) current. The Level-1 (L1) trigger rate
and the occupancy in the other detector systems also constitute occasional
limitations. Careful measurement, analysis and simulation of the background 
sources and their impact have led to a detailed understanding and an 
effective remediation of their effects.

\par
The primary causes of steady-state backgrounds in \pep2\ are, in order of 
increasing importance:
\begin{itemize}
\item Synchrotron radiation generated in the bending magnets and
final focusing quadrupoles in the incoming HER and LER beam lines. Careful
layout of the interaction-region area and a conservative synchrotron radiation
masking scheme have proven very effective against these sources.
\item The interaction of beam particles with residual gas around the
rings (beam-gas), which constitutes the primary source of radiation
damage and has had, averaged over this first run, the largest impact on 
operational efficiency.
\item Collision-related electromagnetic shower debris, dominated by energy-
degraded \epm\ from radiative-Bhabha scattering which strike vacuum
components within a few meters of the interaction point (IP). This background, 
directly proportional to the instantaneous luminosity, was barely detectable 
in early running; it now noticeably affects all detectors except 
the SVT. 

\end{itemize}

While instantaneous background conditions vary because of
interaction point orbit drifts and sensitivity of beam tails to small tune
adjustments, reproducible patterns have emerged. 

The relative importance of the single-beam and luminosity contributions varies 
from one detector system to the next. 
The HER beam-gas contribution significantly 
impacts all detectors except the particle identification system
(DIRC): the combination of a 40~m\ long straight section, almost devoid of 
magnetic bending upstream of the final doublet, and of the magnetic beam-
separation scheme, result in abundant bremsstrahlung-induced electromagnetic 
debris being directed onto the IP vacuum pipe. The same process occurs in the 
LER, but to a lesser extent (for a given beam current) because of a shorter 
drift section and lower primary energy. Most detectors, therefore, 
exhibit occupancy peaks at $\phi = 0$ and 180\degrees, reflecting the fact 
that the separation dipoles bend energy-degraded particles in opposite 
directions. Such local beam-gas interactions contribute noticeably to the 
DCH current; they dominate the SVT instantaneous dose rate in 
the horizontal plane, the EMC crystal occupancy above the cluster-seeding 
threshold (10 MeV), and the L1 trigger rate. Analysis of pass-through events 
from the charged L1 trigger shows that the beam-pipe wall and several aperture 
restrictions within 100\cm\ of the interaction point are the primary impact 
points for lost particles, as illustrated for the HER in 
Figure~\ref{fig:L1vtx} (left)). To minimize this background it is vital to 
maintain a low pressure in the region from 4\m\ to 60\m\ upstream of the 
interaction point in the incoming HER and LER beam lines. Scrubbing, which has 
reduced the average dynamic pressure in both rings to levels close to or below 
the design value, also proved effective (Figure~\ref{fig:L1vtx} (right)).

Under typical running conditions, the DIRC and the DCH receive comparable 
contributions from beam-gas Coulomb scatters around the entire LER, and from 
luminosity backgrounds. In addition, these two detectors proved particularly 
vulnerable to tails generated by beam-beam or electron-cloud induced blowup of 
the low-energy beam. Even though partially eliminated by a set of betatron 
collimators in the last arc, such tails - which are very sensitive to 
accelerator conditions and tuning procedures - tend to scrape near the 
highest-$\beta$ point of the final LER  doublet, located inside the DIRC's 
standoff-box (SOB). The resulting background fluctuations occasionally cause 
the DCH high-voltage protection system to trip, or the counting rate in the 
DIRC photomultipliers to exceed 200~kHz. The problem could be only partially 
alleviated by local lead shielding; additional collimation will be installed 
during the fall 2000 shutdown. 

\begin{figure}[htb]
\begin{center}
\includegraphics[width=3.25in]{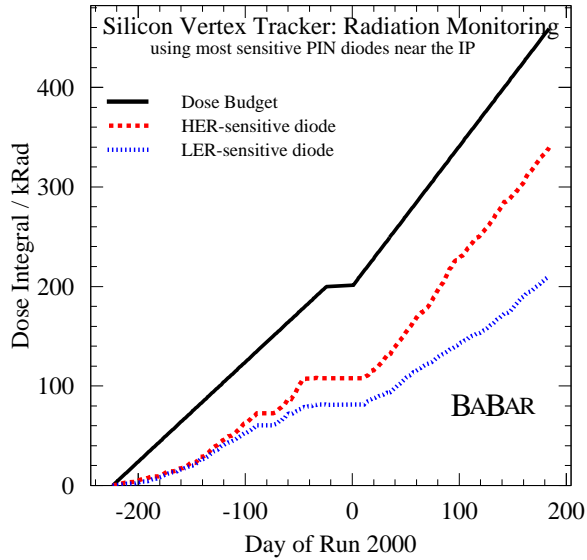}
\end{center}
\caption{ 
projected
History plot of accumulated radiation dose in the SVT using
PIN diodes near the interaction point in the horizontal plane (the black line 
represents the allotted budget, rationing out a 2~MRad dose over
the detector's planned life span).
}
\label{fig:Scrub}
\end{figure}

While trigger-rate and occupancy considerations define
acceptable dynamic running conditions, it is the total integrated
radiation dose that determines the lifetime of the detector systems.
Despite a significant investment in radiation-hard technology, 
the innermost silicon layers of the SVT and its front-end electronics remain 
the most susceptible to radiation damage. The accumulated dose has been 
maintained below budget, as shown in Figure~\ref{fig:Scrub}, through a
strict program of hardware interlocks, administrative controls, and
real-time monitoring. As of this conference, the worst irradiated spot of the 
SVT has been exposed to approximately 300~kRad, 20 to 30\% of which is 
contributed by injection periods.

\renewcommand{\secname}{BaBar}         
\section{The \babar\ Detector}
\label{sec:\secname}

\begin{figure}[!htb]
\begin{center}
\mbox{\epsfig{file=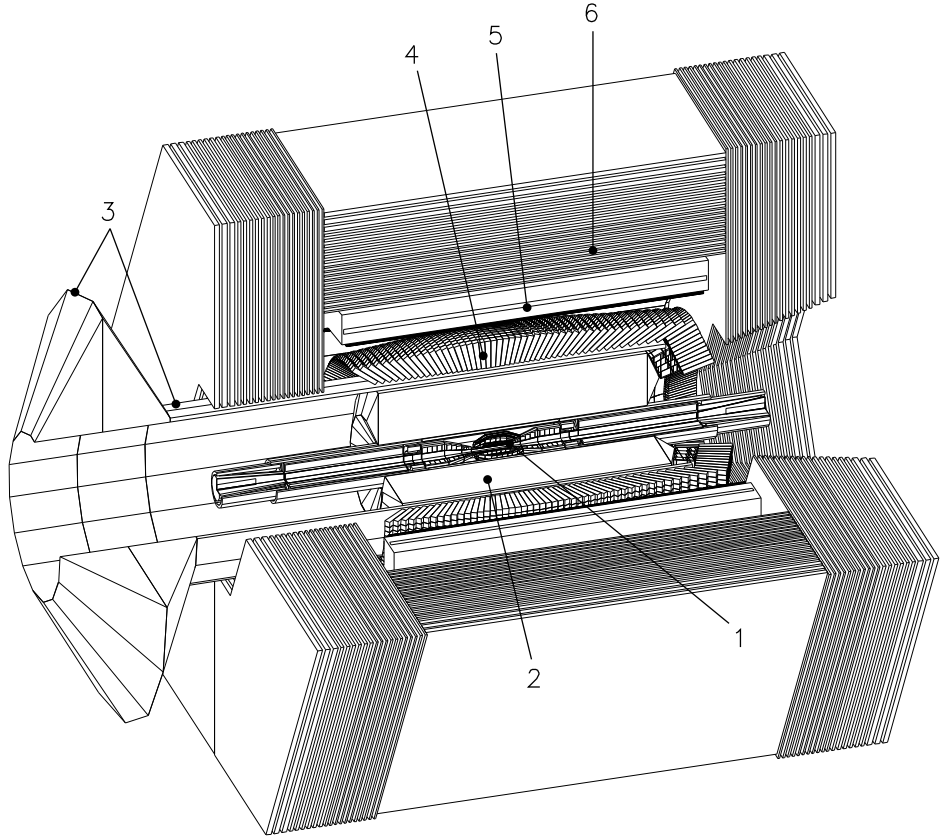,width=0.75\textwidth}}
\end{center}
\caption{
The \babar\ Detector. 1. Silicon Vertex Tracker (SVT), 2. Drift Chamber (DCH),
3. Particle Identification Subsystem (DIRC--Detector of Internally
Reflected Cherenkov Light, 4. Electromagnetic Calorimeter (EMC), 5. Magnet,
6. Instrumented Flux Return (IFR).
}
\label{fig:babarSideCutView}
\end{figure}

 The new magnetic spectrometer \babar\ (Figure~\ref{fig:babarSideCutView}) has 
 been constructed at SLAC, by a
 collaboration of nine countries, to precisely measure \epem\ annihilations at
 center of mass energy $\sqrt{{\rm s}} \sim$10\gev\ produced with the new
 \pep2\ Asymmetric Storage Rings. Construction of this detector was approved in
 November 1995 and its commissioning followed in the Fall of 1998.
 First data with \pep2\ colliding beams was collected in May 1999.


 The \babar\ superconducting solenoid, which produces a 1.5T axial magnetic
 field, contains a set of nested detectors: a five-layer silicon-strip vertex detector (SVT), a
central Drift Chamber (DCH) for charged particle detection and momentum measurement, a quartz-bar
 Cherenkov radiation detector (DIRC) for particle identification, and a CsI
 crystal Electromagnetic Calorimeter (EMC) for detecting photons and separating
 electrons from charged pions. The calorimeter has a barrel section, and an endcap
 which extends it asymmetrically into the forward direction (\en\ beam
 direction), where many of the collision products emerge. (There is no
 calorimeter coverage in the backward direction.)  Two layers of
 cylindrical resistive plate chambers (RPCs) are located between the barrel
 calorimeter and the magnet cryostat. All the detectors located inside the
 magnet have full acceptance in azimuth. The Instrumented Flux Return
(IFR) outside the 
 cryostat is composed of 18 layers of steel, which increase in thickness outward, and contains 19 layers of planar RPCs in the barrel and 18 in the endcaps.
 The RPCs allow the separation of muons and charged hadrons, and also detect
 penetrating neutral hadrons (mainly $K_L$). 

 The following sections provide more details on the subsystems of the detector
 together with current performance results. 

\subsection{Silicon Vertex Tracker}
\label{sec:\secname:SVT}
The Silicon Vertex Tracker (SVT) provides the required vertex resolution 
for the measurement of \CP\ violation
at \pep2. In addition, it is capable of independent charged particle tracking, of particular 
importance for low transverse momentum
particles (\pt\ $<120$\mevc) that cannot be measured by the central tracking chamber. 
The detector design is optimized to take into account the physical
constraints imposed by the \pep2\ geometry at the interaction region: the
presence of the B1 permanent magnets at $\rm{\pm 20}$\cm\ from the interaction
point, which separate the beams after head-on collisions. 
The acceptance in polar angle $\theta$ is limited
by the gap between beamline elements to $-0.87 < \cos  \theta_{lab} < 0.96$
($-0.95 < \cos  \theta_{cm} < 0.87$). 
As the innermost \babar\ subdetector, two important considerations in
optimizing the design were low mass, to minimize multiple
scattering, and radiation hardness of its components. 
A detailed description of the SVT and
its components can be found in Ref.~\cite{SVT}.

\begin{figure}[!htb]
\begin{center}
\mbox{\epsfig{file=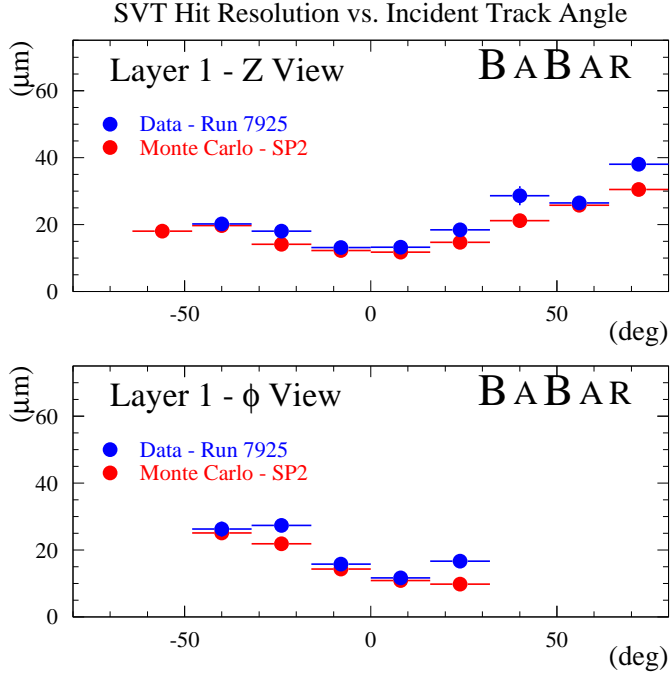,width=0.6\textwidth}}
\end{center}
\caption{SVT single point resolution as a function of the 
incident track angle for the innermost layer.}
\label{fig:svt_resol}
\end{figure}

The SVT contains 52 modules built out of AC-coupled double-sided silicon sensors
(strips othogonal on the two sides). These are read out by a full-custom
low-noise
radiation-hard integrated circuit, known as the AtoM chip (mounted on a passive
hybrid circuit that is attached to a water cooling channel). The detector modules are
organized in five radial layers, each divided azimuthally into 6, 6, 6, 16 and 18
staves respectively (see Table~\ref{tab:svt_geo}). The three inner layers are 
crucial for vertex and
tracking resolution, while the outer two layers are needed to provide additional
measurements for stand alone tracking independent of drift chamber
information. These outer two layers each contain two different types of modules,
an inner (labelled $a$ in the Table~\ref{tab:svt_geo}) and an outer (labelled $b$) layer,
occupying slightly different radial positions.
The modules are assembled on
carbon fiber support cones, which in turn are positioned around the beam pipe
and the B1 magnets.
The SVT and some beamline elements are  housed
inside a strong support tube, with its load transferred at the ends to the
\pep2\ beamline support ``rafts.''

\begin{table}[!htb]
\caption{Layer Structure of the SVT.}
\begin{center}
\begin{tabular}{@{}crccrr@{}}\hline
Layer & Radius & Modules & Si Wafers & $\phi$ pitch & $z$ pitch \\
      & (mm)   & /Layer  & /Module   &  ($\mu$m)     & ($\mu$m) \\ \hline
1     &  32    &   6     & 4         &     50 or 100 & 100      \\
2     &  40    &   6     & 4         &     55 or 110 & 100      \\
3     &  54    &   6     & 6         &     55 or 110 & 100      \\
4a    & 124    &   8     & 7         &    100        & 210      \\
4b    & 127    &   8     & 7         &    100        & 210      \\
5a    & 140    &   9     & 8         &    100        & 210      \\ 
5b    & 144    &   9     & 8         &    100        & 210      \\ \hline
\end{tabular}
\end{center}
\label{tab:svt_geo}
\end{table}

A system consisting of 12 PIN photodiode sensors is placed close to the
first silicon layer to monitor continuously the radiation exposure of
the SVT and manage its radiation dose budget.

\begin{figure}[!htb]
\begin{center}
\mbox{\epsfig{file=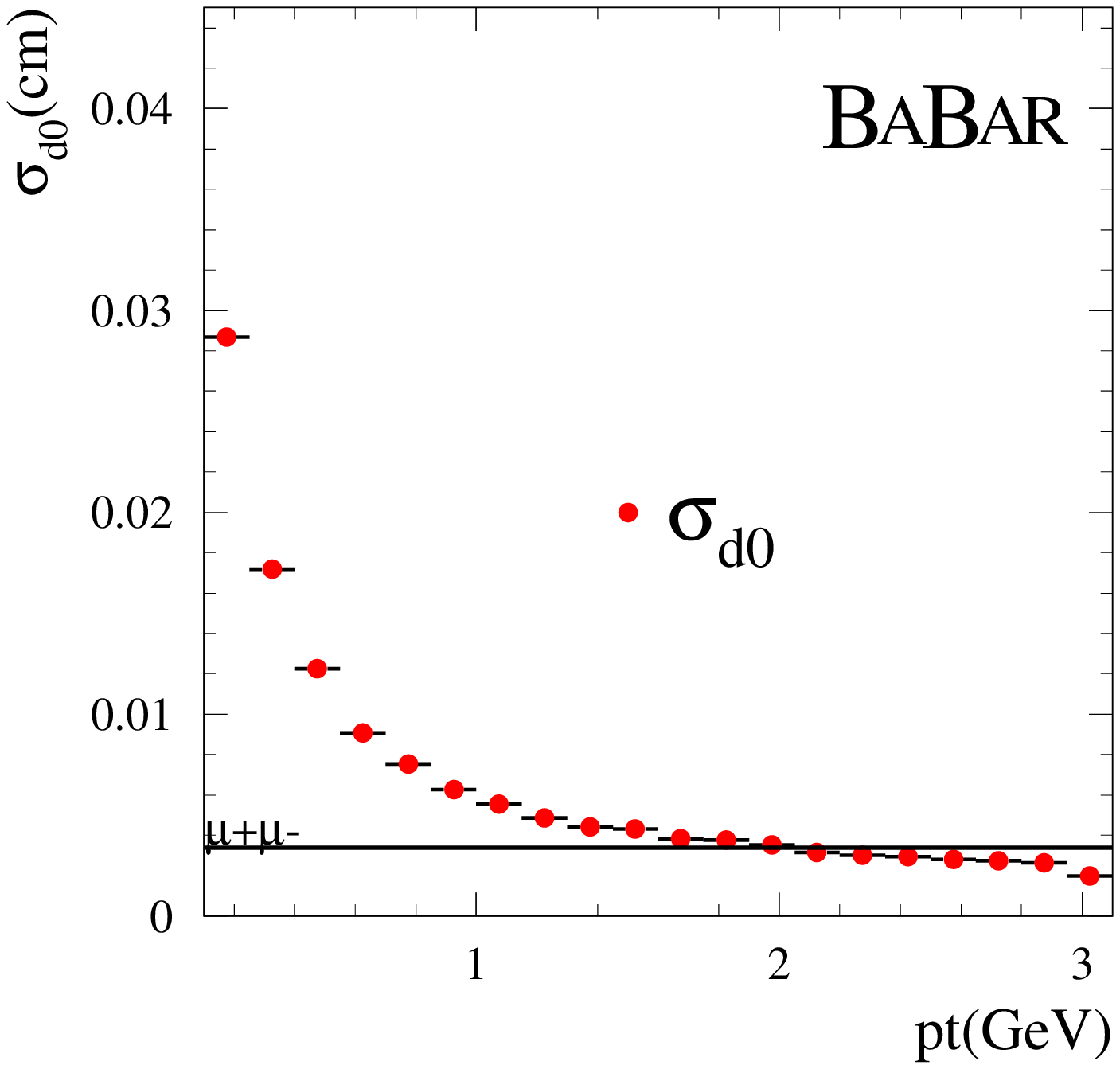,width=0.45\textwidth}}
\mbox{\epsfig{file=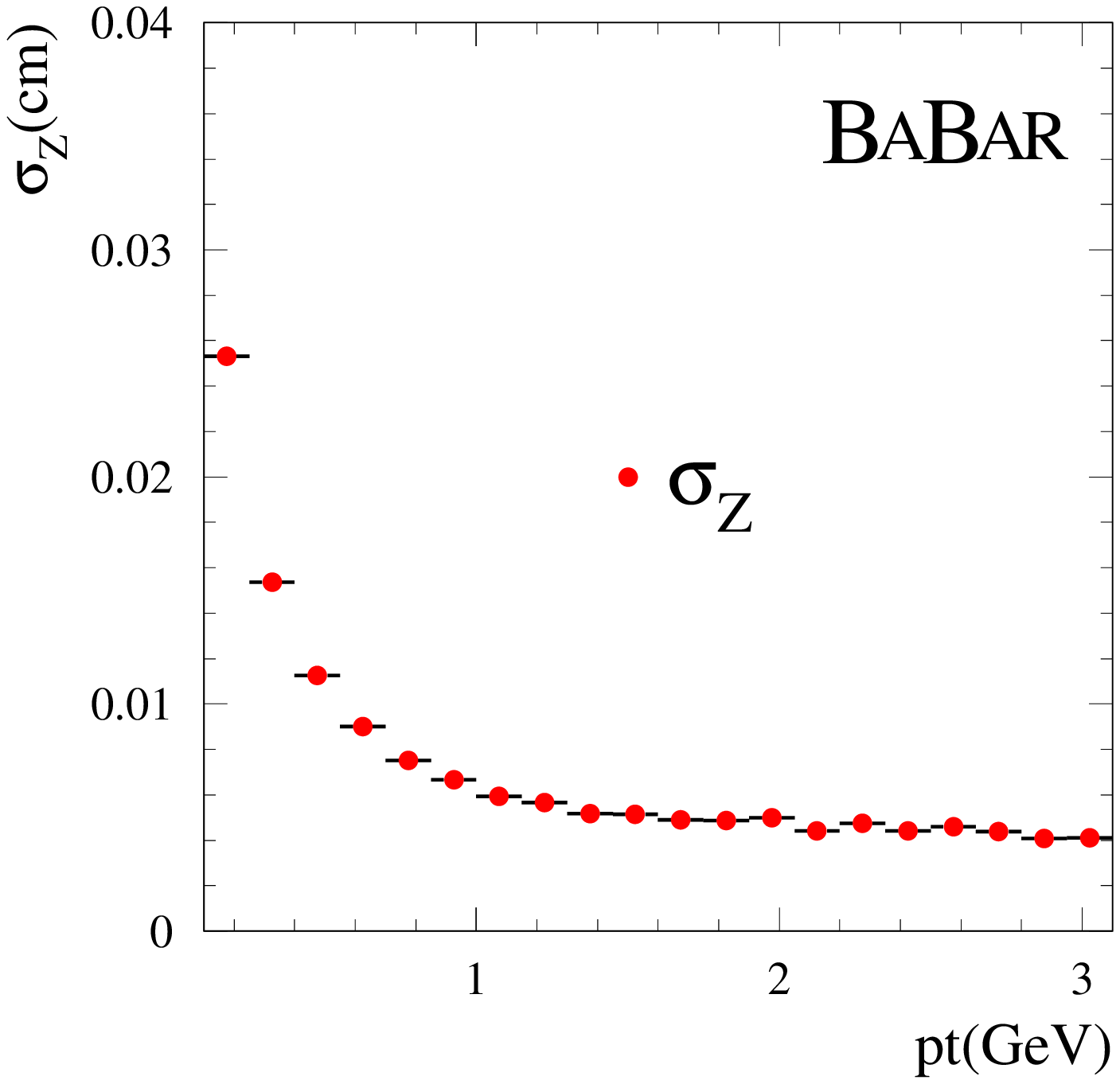,width=0.45\textwidth}}
\end{center}
\caption{Impact parameter resolution in the $r-\phi$ ($\sigma_{d0}$ in the left plot) and in the
$z$-view ($\sigma_z$ in the right plot) as measured on data. 
The horizontal line in the left-hand plot is $1/\sqrt{2}$
times the transverse miss distance for muon pair events.}
\label{fig:svt_imp_param}
\end{figure}

During the first year of data taking, all
major design goals for the detector
were already achieved. The average hit reconstruction efficiency is above 98\% in
both views. The hit resolution presently observed, following an 
initial accurate alignment
procedure, is shown in Figure~\ref{fig:svt_resol}. This compares
well with Monte Carlo predictions for the resolution with perfect alignment. The impact parameter
resolution, dominated by the SVT's precision for
measuring high transverse momentum
tracks, is shown in Figure~\ref{fig:svt_imp_param}. Two-track vertices, such as  $J/\psi \rightarrow \mu^+ \mu^-$ are
reconstructed with a typical resolution of 50\mum; the resolution on the z
separation between the two \B\ decay vertices is typically 110\mum, in good
agreement with the design goals.

\subsection{Drift Chamber}
\label{sec:\secname:DCH}
The Drift Chamber (DCH) is the main tracking device for charged particles 
with transverse momenta \pt\ above about 120\mevc, providing  a precision measurement of \pt\ from the curvature of charged 
particle trajectories
in the 1.5$\,{\rm T}$ magnetic field. Prompt single-cell hit information
from  the chamber
is also a basic component of the  Level-1 trigger (see Section 3.6.1). The DCH is a 280\cm-long cylinder, with a inner radius of 
23.6\cm\ and and outer radius of 80.9\cm. It is mounted in cantilever from the rear endplate to the 
DIRC central support 
tube, within the volume inside the DIRC and outside the \pep2\ support tube. The 
center of the chamber is displaced forward by 36.7\cm\ to improve the forward track length, given the
asymmetric boost for the \FourS\ events. The beryllium inner wall (0.28\% radiation lengths), the thin outer
half of the forward endplate (15\cm\ aluminum), and the carbon-fiber outer cylinder are all designed
to minimize material in front of the calorimeter. 

The drift system  consists
of 7104 hexagonal cells, approximately 1.8\cm\ wide by 1.2\cm\ high,
arranged in 40 concentric layers between a radius of 25.3 and 79.0\cm. 
The active volume provides charged particle tracking over the polar angle range $-0.92 < \cos
\theta_{lab} < 0.96$. The forty layers are grouped
into ten superlayers of four layers each, organized with the same orientation for sense and field wires
within a given superlayer. The superlayer structure facilitates fast local segment finding as the
first step in pattern recognition. This arrangement is particularly important for Level-1 trigger decisions. 
Superlayers alternate in orientation: first axial (A), then  a small positive 
stereo angle (U), followed by a small negative stereo angle (V). All superlayers participate in the Level-1 trigger track finding; only
the axial superlayers participate in the Level-1 trigger \pt\ determination.

\begin{table}[!htb]
\caption{Layer arrangement for the DCH}
\vspace{0.cm}
\begin{center}
\label{tab:dch}
\begin{tabular}{|c|c|c|c|} \hline
Superlayer & Inner       & Cells/      & Stereo           \\
           & Radius [cm] & Layer       & Angle [mr]     \\ \hline\hline
1          & 26.04       &  96         & 0                \\
2          & 31.85       &  112        & $+$[44.9--50.0]  \\
3          & 37.05       &  128        & $-$[52.3--57.4]  \\
4          & 42.27       &  144        & 0                \\
5          & 48.08       &  176        & $+$[55.6--59.7]  \\
6          & 53.32       &  192        & $-$[62.8--66.9]  \\
7          & 58.54       &  208        & 0                \\
8          & 64.30       &  224        & $+$[65.0--68.5]  \\
9          & 69.52       &  240        & $-$[72.1--75.8]  \\
10         & 74.72       &  256        & 0                \\
\hline
\end{tabular}
\end{center}
\end{table}

The individual hexagonal  cells each consist of a 20\mum\ 
rhenium-tungsten sense wire operating
nominally in the range 1900--1960${\,\rm V}$, surrounded by 6 cathode wires,
approximately half of which are shared with
neighboring cells. For the inner cells in a superlayer, the cathode wires
are  120\mum\  aluminum, grounded to the rear endplate. For outer cells one
of the cathode wires is  80\mum\ aluminum, held  at about 350$\,{\rm V}$.
All sense and cathode wires are gold plated.
Table~\ref{tab:dch} summarizes 
the main geometric features of the design.
The counting gas, chosen to have low mass, is  an 80\%:20\% mixture of He:Isobutane, with a small amount 
(3000~ppm) of water vapor to prolong
the lifetime of the chamber in a high rate environment. The combination of low-mass gas and minimization
of material in the field cage itself is designed to reduce the contribution of multiple scattering to
the \pt\ resolution for the typically soft tracks produced in \FourS\ events.

The DCH readout electronics, including front-end amplifier, timing and pulse height digitization, 
event readout and high voltage distribution, are all mounted on the rear endplate 
(opposite the boost direction) to keep the additional material outside the detector fiducial volume. 
The instrumentation includes two custom integrated circuits: a low-noise bipolar amplifier IC and 
a CMOS digitizer IC that incorporates 8 channels of TDC, Flash-ADC, pipelined data-storage, multiple 
event buffers and prompt trigger output.  Data readout is multiplexed on four fiber-optic links 
to \babar-standard VME readout modules (ROM);  trigger data is transported on 24 fiber-optic 
links. The ROMs extract the integrated charge from the digitized waveform before passing the 
data on to next level in the data acquisition system.  High voltage is supplied by a CAEN SY527 
mainframe and is distributed through circuit boards that plug directly on the DCH 
endplate feedthroughs.  Temperature and humidity sensors monitor the chamber
environment, and  radiation monitors  track the accumulated dose. 

\begin{figure}[!htb]
\begin{center}
\begin{tabular}{lr}
\mbox{\epsfxsize=3.in\epsffile{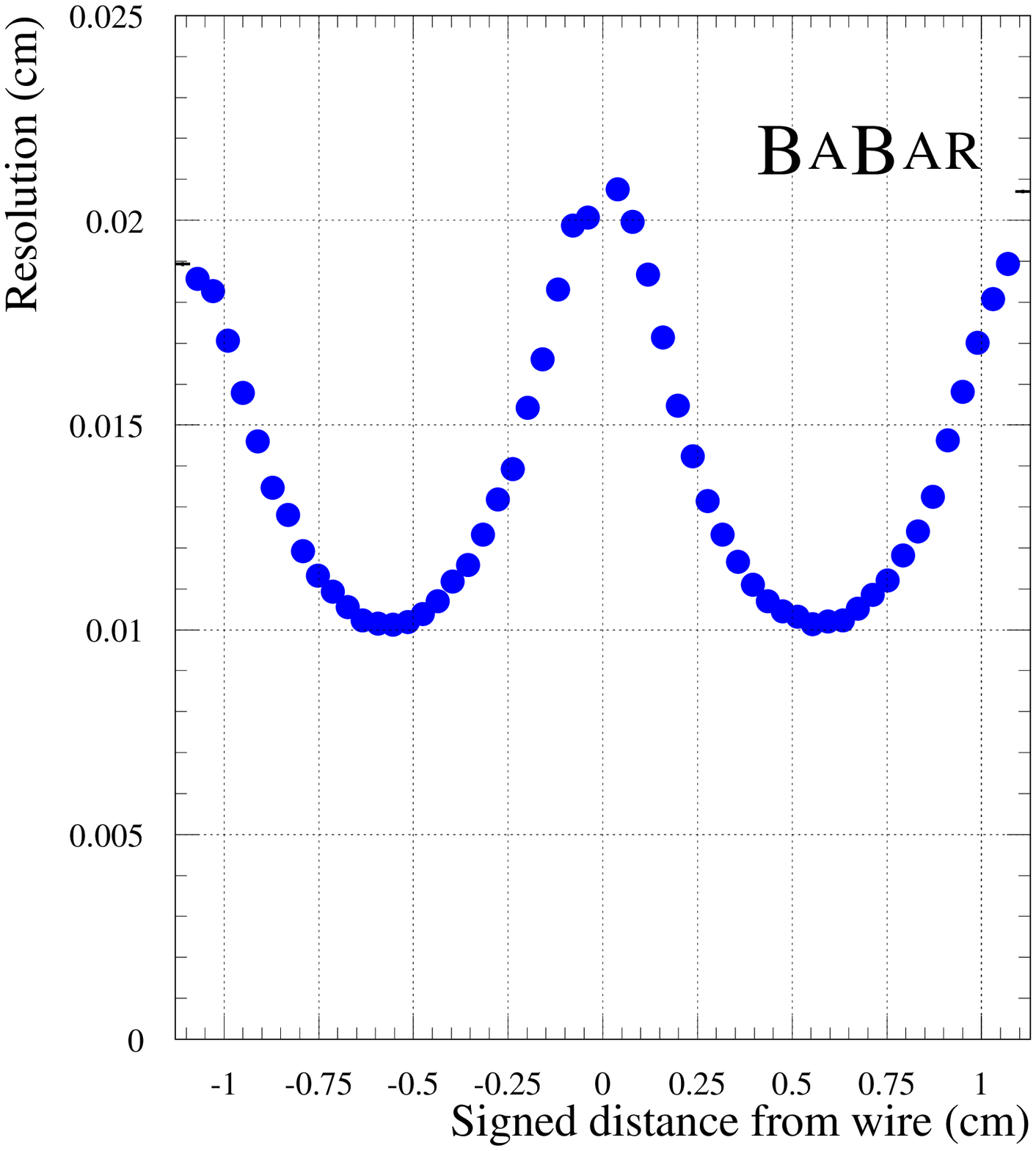}} &
\mbox{\epsfxsize=3.in\epsffile{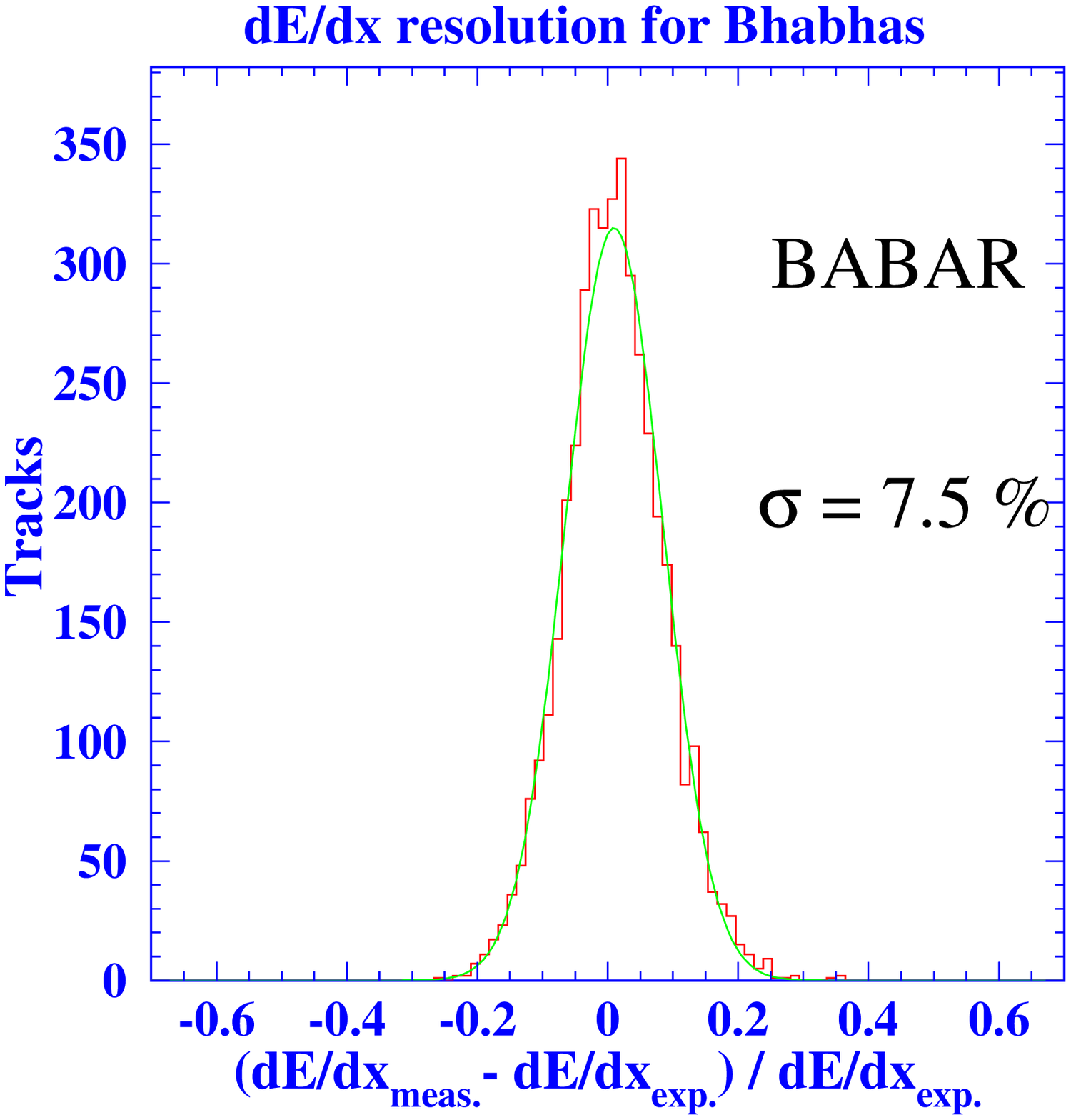}} \\
\end{tabular}
\end{center}
\caption{
DCH single cell resolution (left) and dE/dx resolution for Bhabha electrons (right).}
\label{fig:dchres}
\end{figure}

The nominal design for the chamber calls for an average 140\mum\ single point 
resolution. The space-time relation (STR) for the non-saturated counting gas has been 
modeled with a separate 6$^{th}$-order Chebychev polynomial for the left and right 
parts of the cell. The STR has proven to be very stable over time. There is a small
residual dependence on gas density, not yet corrected in production. The single
cell resolution function obtained from the ensemble of charged tracks in the
normal event stream is shown in Figure~\ref{fig:dchres} for operation at 1960$\,{\rm V}$.
This shows a weighted average resolution of 125\mum.

In addition to recording timing information for hits, 
the Flash-ADC is used to measure the time development of pulse height in the cell.
Signal processing (feature extraction) in the ROM  converts this into a measurement of deposited energy.
Real-time overall gain corrections are calculated and applied during the
reconstruction step.
Saturation, path length, polar angle, individual wire and layer gain
corrections are applied as well,
although these corrections are stable over long periods of time. The resulting distribution of
the truncated mean dE/dx observed for Bhabha electrons is shown in Figure~\ref{fig:dchres}.
The observed resolution is found to be 7.5\%. We expect to achieve the predicted
resolution of 7\% after additional corrections, including the effect of varying entrance angles.

\subsection{Particle Identification System}
\label{sec:\secname:DRC}
Standalone identification of charged particles
is based on a specialized subdetector system that uses 
the Detection of Internally Reflected Cherenkov light
(DIRC). Charged particles exiting the
barrel region of the DCH transit an array of 144 fused silica quartz
bars, each approximately 17\mm\ thick ($\delta r$), 35\mm\ wide ($\delta(r\phi)
$) and 4.9\m\ long ($\delta z$). Particles above Cherenkov threshold
radiate photons in the quartz media. The angles of the Cherenkov photons with
respect to particle direction are measured with an array of 10,752
photomultiplier tubes located outside the return yoke, in a special
low magnetic field volume. The polar angle coverage is $-0.84 < \cos \theta_{lab}
< 0.90$.
\par
\begin{figure}[!htb]
\begin{center}
\mbox{\epsfig{file=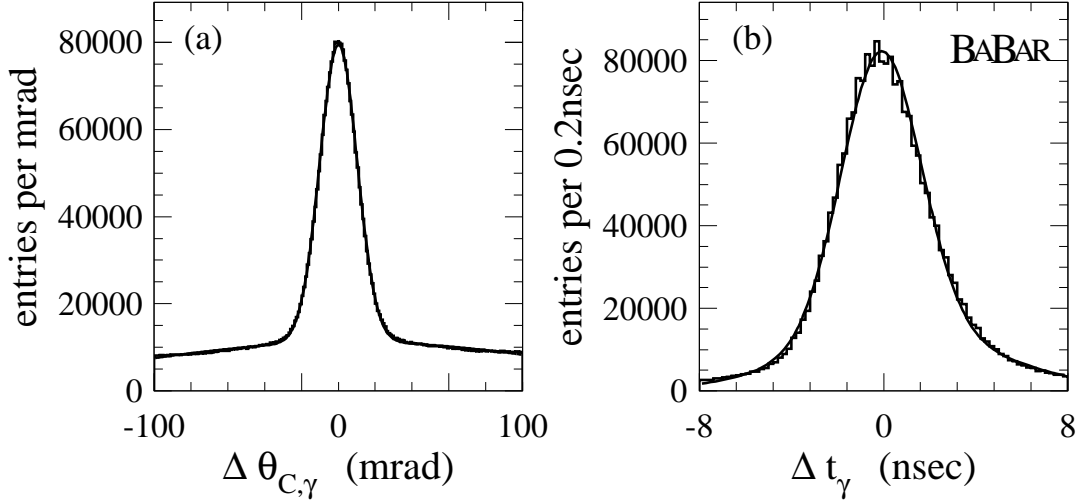,width=0.95\textwidth}}
\end{center}
\caption{Resolution of the reconstructed Cherenkov angle for single photons
(a) and of the difference between measured and expected arrival time (b).}
\label{fig:DIRC}
\end{figure}

The 144 quartz bars of the DIRC are arranged in 12 modules, or ``bar-boxes'', that penetrate
through the magnetic end plug. Cherenkov photons trapped in a quartz
bar exit the bar through a wedge and a quartz window into a water tank, which optically
couples the quartz bars to the photomultiplier array. The photomultiplier tubes
are arranged in 12 sectors corresponding to the 12 barboxes, on the
surface of a half torus, with a major radius of about 3\m\ and a minor radius
of 1.2\m. By knowing the particle direction from the tracking system, and the
location of the photomultiplier tube that observes a Cherenkov photon, the angle
that a Cherenkov photon makes with respect to the track direction can be 
deduced. Due to internal reflections within the bars, there
are several solutions for a hit-to-track association. 
The angular resolution for a single photon is about 10.2\mrad\ (see
Figure~\ref{fig:DIRC}(a)) 
and, with an 
average of
30 photons per track, the ``per track'' Cherenkov resolution is about 2.8\mrad, rms.
This corresponds to a separation of approximately three standard deviations
between charged  $K$'s and  $\pi$'s at 3\gevc. An additional observable is the time of a
photomultiplier hit, measured to a precision of 1.7\ns\ (see
Figure~\ref{fig:DIRC}(b)). By comparison of the measured hit time with the
propagation time of a photon solution, it is possible to effectively suppress
background photons and invalid photon solutions.  
  
\subsection{Electromagnetic Calorimeter}
\label{sec:\secname:EMC}
The Electromagnetic Calorimeter (EMC) contains 6580 CsI crystals doped with
thallium iodide at about 1000~ppm. Each crystal is a truncated trapezoidal
pyramid and ranges from 16 to 17.5 radiation lengths in thickness. The front
faces are typically $\sim$5\cm\ in each dimension. The crystals are arranged
quasi-projectively in a barrel structure of 48 polar ($\theta$) rows by 120 crystals in
azimuth ($\phi$), with an inner radius of 90\cm. The forward end is closed by a
separable endcap capable of holding nine additional rows. At present, eight are
filled with crystals and the innermost with shielding absorber. The polar angle
coverage of the calorimeter is $-0.78 < \cos \theta_{lab} < 0.96$. Beamline
elements occlude $\cos\theta_{lab} > 0.94$. Each crystal is wrapped with a
 diffuse reflective material
(TYVEK) and housed in a thin eggcrate-like carbon fiber composite mechanical
structure. There are 280 such modules in the barrel (7 types, 40 of each type)
 and 20 identical endcap modules. Crystals are read out with two independent
 2\cma\ large area PIN photodiodes epoxied to their rear faces. Dual-range ASIC
 preamplifiers reside directly behind the photodiodes in a shielded housing that
 also provides a thermal path for heat removal. Shielded ribbon cables carry
 analog signals to the end flanges of the barrel and the back plate of the
 endcap, where additional amplification and digitizing electronics 
provide for a total of four overlapping linear ranges. The system handles
signals from $\sim$50\kev\ to $\sim$13\gev, corresponding to 18 bit dynamic 
range. 
A short shaping time of $\sim 400\ns$ 
 is used in the preamplifiers to reduce the impact of soft ($<$5\mev) beam-related 
 photon backgrounds. Noise performance can be recovered by digitally
 processing the signal waveform sampled at 4\,MHz. Calibration and monitoring is achieved
 by charge injection into the front end of the preamplifiers, 
a fiberoptic/xenon
 pulser system injecting light into the rear of each crystal, and a circulating
 radioactive source (a neutron-activated fluorocarbon fluid) producing a 6.13\mev\
 photon peak in each crystal. Signals from data (\piz s, $\eta$s, radiative, 
and non-radiative
  Bhabhas, $\gamma\gamma$ and \mumu\ events) can provide additional calibration
  points. Source and Bhabha calibrations are updated weekly to track the small
  changes in light yield with integrated radiation dose. Light pulser runs are
  carried out daily to monitor relative changes at the $<$0.3\% level.

The calorimeter has achieved an electronics noise energy (ENE)
of $\sim$220\kev\ (coherent plus incoherent) measured with the source system 
(in the absence of colliding beams) after digital signal processing. 
During regular data taking, this digital filtering is not applied and the 
ENE rises to $\sim$450\kev\ owing to the short shaping time; consequently,
only channels with $>$1\mev\ are presently used in the reconstruction
of calorimeter energy deposits.
The efficiency of the calorimeter exceeds 96\% for the detection of 
photons with energies above 20\mev. 
  


The energy resolution can be measured directly with the radioactive source at 
low energies and with
electrons from Bhabha scattering at high energies, yielding
resolutions of $\sigma \rm{(E)/E} = 5.0 \pm 0.8\%$ at 6.13\mev\
and $\sigma \rm{(E)/E} = 1.9 \pm 0.07\%$ at 7.5\gev, respectively. 
The energy resolution can also be inferred from the observed mass
resolutions for the \piz\ and $\eta$, which are measured 
to be around 7\mev\ and 16\mev, respectively.

\begin{figure}[!htb]
\begin{center}
\mbox{\epsfig{file=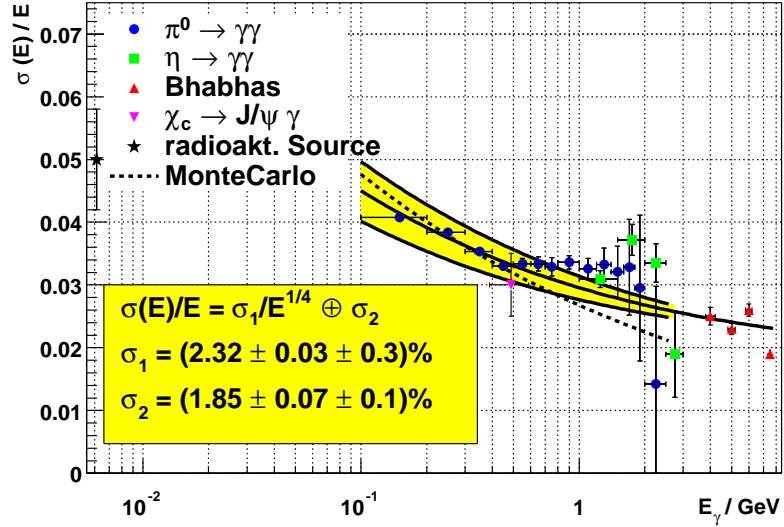,width=0.7\textwidth}}
\end{center}
\caption{The energy resolution as a function of energy, as determined from the 
observed width of \piz\ and $\eta$ decays to two photons of equal energy, 
and the resolution for Bhabha electrons. 
The shaded band is the best fit to 
the \piz, $\eta$, and Bhabha data. Also shown is the energy resolution of the 
6.13\,MeV photons from the radioactive source, and of the photons in the 
transition $\chic{1}\ra\jpsi\gamma$.}
\label{fig:EMC_eres}
\end{figure}

Figure~\ref{fig:EMC_eres} shows the energy resolution extracted from a 
variety of data as a function of energy.
A fit to the \piz, $\eta$, and Bhabha energy resolution measurements, 
assuming an energy dependence of the form:
\begin{equation} 
\sigma \rm{(E)/E} = \sigma_1 \rm{(E/GeV)}^{-1/4} \oplus \sigma_2 ,
\end{equation}
gives
$\sigma_1 = ( 2.32 \pm 0.30 )\%$ and 
$\sigma_2 = ( 1.85 \pm 0.12 )\%$.
Implementation of digital filtering will reduce the electronics and beam 
background noise. This, together with the afforded improvements in cluster 
reconstruction algorithms, will lead to better energy resolution for the 2001 
run, particularly at lower energies.

The \piz\ and $\eta$ data are also used to measure the angular resolution of
the calorimeter. It is found to vary between about 12\mrad\ at low energies 
and 3\mrad\ at high energies, described by an energy dependence according to:
\begin{equation} 
\sigma_{\theta,\phi} = \sigma_1 \rm{(E/GeV)}^{-1/2} + \sigma_2 ,
\end{equation}
with
$\sigma_1 = ( 3.87 \pm 0.07 )$\,mrad and 
$\sigma_2 = ( 0.00 \pm 0.04 )$\,mrad.

The calorimeter is also used for the separation of hadrons from electrons, 
and in conjunction with the IFR for muon and \KL\ identification;
its performance in these areas is presented in section 7.2.

\subsection{Instrumented Flux Return}
\label{sec:\secname:IFR}
The Instrumented Flux Return (IFR) is used to identify muons and neutral
hadrons.  The detector consists of nearly 900 Resistive Plate Chambers (RPCs)
interleaved with the iron plates which comprise  the flux return for the 1.5~T
solenoidal magnet. There are 19 RPC  layers in the barrel
region and 18 layers in the forward and backward endcaps. The iron plate
thickness is graded from 2\cm\ for plates closest to the interaction region to
10\cm\ for the outermost layers for a total of $\geq$65\cm\ in the barrel and
$\geq$60\cm\ in the endcaps. An additional cylindrical RPC containing two layers
is located in the barrel region between the calorimeter and solenoid cryostat.

The single gap planar RPC's  ~\cite{Santonico} are constructed from 2 sheets of bakelite
separated by 2\mm\ spacers. The inner surfaces are treated with linseed oil. The
outer bakelite surfaces are painted with graphite. Layers of insulating mylar,
aluminized mylar pickup strips, foam and an aluminized mylar ground-plane are
glued to both sides of the bakelite forming the RPC  package. The pickup strips
are connected with discriminator cards which record a bit per strip if an
in-time signal above threshold is recorded. The pickup strips on top and bottom
are orthogonal, providing x-y position information in the endcaps and
$\rm{\phi}$-z information in the barrel. The cylindrical RPC's
 are constructed
from a special plastic requiring no surface treatment.
The RPC's  utilize a gas mixture of
56.7\% Argon, 38.8\% Freon-134a, and 4.5\% Isobutane and operate at 7.6~kV. The
cylindrical RPC's use the same gas but run at 7.4~kV. The
steel flux return is water-cooled to reduce temperature excursions, keeping  the
barrel at $\sim$20\degc\ and the endcaps at $\sim$22--24\degc.


The RPC  efficiencies are measured in cosmic ray runs taken weekly and in
collision data. The average chamber efficiencies during the 2000 run were
$\sim$78\% in the barrel, and $\sim$87\% in the forward endcap, lower than
the average $\sim$92\% measured in June 1999. A small fraction of the RPC's
is presently disconnected because of high-voltage problems. A vigorous
testing and R\&D program is underway to understand the efficiency decrease
and to improve RPC performance.

  
\subsection{Trigger}
\label{sec:\secname:Trigger}

The \babar\ trigger system consists of the Level-1 hardware trigger
and the Level-3 software trigger. The Level-1 trigger decision is issued
within a latency of 11--12~$\mu$s after the corresponding
beam crossing to initiate the Data Acquisition (DAQ) readout of the
relevant time slice of the pipelined data for all detectors.
The latency for hadronic events is confined to a smaller, $\pm 150$ns,
range.
The acquired data
are processed by Level-3 running on the online farm nodes to select physics
and calibration events for data logging.

\subsubsection{Level-1 Trigger}

The Level-1 (L1) trigger system comprises the DCH track trigger (DCT), the
EMC energy trigger (EMT), the IFR muon trigger (IFT) and the
global trigger (GLT). The L1 trigger system operates in a continuous 
sampling
mode, processing input data and generating output trigger
information at fixed time intervals. The main logic of the L1 trigger
is implemented in six types of 9u VME modules with a total of 41 boards. 
The basic trigger primitives generated by the DCT, EMT and IFT, mostly
$\phi$ maps of the particle signatures, are collected by the GLT to
form 24 trigger lines for the fast control system.
The DCT identifies short and long track
primitives as well as high \pt\ track primitives. Using the loosest
criteria,  short tracks are accepted down to  $\pt > 120$\mevc\ if they reach
 superlayer five in the DCH. The EMT generates trigger primitives
passing various energy thresholds. The lowest threshold is 100\mev, 
in order to be fully efficient for minimum ionizing
particles.

All the L1 trigger system components appear to the data acquisition
system as standard detector subsystems, in addition to their role in 
providing the trigger signals.  The data read out from them provides
details on the operation of the trigger for each acquired event, 
facilitating diagnostics and the determination of the L1 trigger 
efficiency.

The L1 trigger system is designed to be able to trigger independently from
pure DCT and pure EMT triggers with high efficiency for most physics 
sources, in order to allow cross-calibration of efficiencies. In
particular, \BB\ events are triggered at $>99\%$
efficiency from either DCT or EMT and the combined efficiency is
$>99.9\%$. Only $\tau$ and two-photon events do not have a fully 
efficient pure EMT trigger and rely mainly on DCT
triggers.

The total L1 trigger rate for a typical run with HER (LER) currents
at 700~mA (1100~mA) and a luminosity of 
2.0$\times 10^{33} \rm{cm^{-2}s^{-1}}$ is $\sim$700~Hz.
 This is typically stable to within 10\% for the same machine
currents but can be 50\% higher after a major shutdown when the vacuum
is relatively poor. The joint trigger/data acquisition system rate capability is
comfortably beyond the design rate of 2~kHz. This avoids significant
deadtime, and leaves room for the expected increase in
luminosity. Within the typical L1 trigger rate of 700~Hz 
at 2$\times 10^{33}\rm{cm^{-2}s^{-1}}$,
Bhabha events and other $e^+e^-$ interactions amount to
around 120~Hz, and cosmic ray interactions to 130~Hz.  
The dominant source of background causing the remaining triggers is 
the interaction of lost particles with the beam line components. 
The L1 background rate coming from the HER  beam is $\sim$3 times
higher than that from the LER beam 
for the typical operating currents.

\subsubsection{Level-3 Trigger}

The Level-3 (L3) trigger is the first component of the trigger system to see
complete events. It is embedded in the Online Event Processing (OEP)
framework running in parallel on 32 event filter farm nodes.  The L3
trigger processes data from the DCH track trigger and from the
EMC using two
independent algorithms that form track and cluster objects.  
These are fed into a set of filter algorithms that
operate on either one or both.  The L3 DCH algorithm performs fast
lookup-table based track finding and three-dimensional track fitting,
efficient for tracks with $\pt> 250$\mevc. The  {\it event} 
$t_0$ is
determined from DCH trigger track segments and hit data to better 
than 5\ns.  The L3 EMC algorithm
uses calorimeter crystal data in a fast two-dimensional clustering
algorithm implemented using a lookup table. To reduce noise, the
crystal energy threshold is set to 30\mev. L3 clusters are accepted
for energies $>100$\mev.

The L3 logging decision is based on generic track/cluster topologies
rather than on the identification of individual physics processes.
An exception is made for Bhabha events, which have to be vetoed to reduce
their rate ($\sim 100$~Hz at $2\times 10^{33}\rm{cm^{-2}s^{-1}}$ luminosity).
The physics trigger is a logical OR of two orthogonal filters.
The track filter requires
either one track with $\pt >800$\mevc\ coming from the interaction point,
or two tracks with $\pt >250$\mevc\ and slightly looser vertex
cuts.  The cluster filter accepts events with a high multiplicity 
or with a
large total energy in the EMC and a high invariant (pseudo) mass.
The efficiency of the track filter for \BB\ events is
$99\%$, while the efficiency of the cluster filter is $94\%$.

Both filters are subject to a veto algorithm that identifies Bhabha
events based on clean signatures in the DCH 
and the EMC. The veto has no
impact on hadronic events, and only has noticeable effect on very
few types of events such as 
$\tau\tau\rightarrow \electron\nu\nu \electron\nu\nu$.
A prescaled sample of
Bhabha events, flattened in $\theta$, is preserved for calibration
purposes.  In addition, L3 logs various other samples for calibration
and monitoring, such as radiative Bhabha events, cosmics, and random
triggers. 
The luminosity is measured online by L3 using a track-based
algorithm with a scale calibration precision of $<\sim 5$\%.  The typical
logging rate at $2.0\times 10^{33} \rm{cm^{-2}s^{-1}}$ is $\sim 85$~Hz,
well within
the design specifications.

\subsection{Data Acquisition and Online System}
\label{sec:\secname:Data Acquisition and Online System}
The \babar\ data acquisition system employs a set of standard Readout Modules 
(ROMs) to communicate with the front-end electronics of the detector systems.
A common communications fabric---1~Gbps optical fiber links---and protocol 
is used for readout by all the systems, with detector-specific extensions to
the protocol for control.

Each ROM is built around a commercial VME single-board computer
running VxWorks, linked with three custom boards, one of which
supplies one of two ``personalities''.
The calorimeter uses an untriggered version of the ROM which continuously 
collects data from up to three input data fibers, separating out in software
the data corresponding to L1 triggers.
All of the other detector systems use a
version which requests and collects event data from up to two fibers upon the
receipt of a L1 trigger.

ROMs are organized into crates in groups of two to eleven, including a master
ROM. 
The master builds partial events from the data acquired by the crate's other 
ROMs from the front ends.
Final event building is performed by sending these data fragments via
100--Mbps switched Ethernet to the nodes of the Online Event
Processing (OEP) farm, which are commercial Unix workstations.  
Static load-balancing is used, allocating events to nodes using a
simple deterministic algorithm based on a unique time stamp assigned
to each L1 trigger.

The farm nodes in OEP apply the L3 trigger algorithms and perform
fast data quality monitoring functions.  
Events passing the L3 selection are sent via TCP/IP to a single
server process which logs them to disk files.
Data quality 
monitoring results are collected from all nodes and merged for display
to operators and for automated comparison against defined references.

This system has easily exceeded its fundamental performance
requirement of acquiring 2000~Hz of L1 triggers and logging
L3 triggers at up to 100~Hz.
In recent running we have been operating far from its limits, with a
L1 rate of approximately 700~Hz, an average event size of 27~kB,
and a L3 logging rate of 85~Hz.  
There is therefore considerable headroom already available for
operation at higher luminosities, but the system can be scaled
further, if necessary, up to the data bandwidth limits of the
front-end electronics through the addition of more ROMs and more OEP
nodes.

The online computing system includes several additional components.
The Detector Control
system, which monitors and controls the experiments environmental 
systems such as low and high voltage power and detector gas supplies,
as well as monitoring parameters of the PEP-II collider.  
The Run Control subsystem coordinates the activities of all the
online and data acquisition components, and interlocks data acquisition
with the maintenance of safe and acceptable detector and collider
conditions.

\subsection{Online Prompt Reconstruction}
\label{sec:\secname:OPR}

The first stage of production processing after data acquisition, 
the Online Prompt Reconstruction or OPR, is performed automatically 
by a compute farm which currently consists of about 150 Unix processors.
All colliding beam events are filtered and tagged. Interesting events 
are completely reconstructed and written to the object database.
A novel system of ``rolling calibrations'' extracts updated 
reconstruction constants from the data which are then written 
to the object database and used for subsequent reconstruction. 
Finally, detailed monitoring distributions are generated and fed 
back into the experiments data quality monitoring program.

OPR is designed and sized to keep up with the data acquisition with 
minimum latency. Typically, this latency between data availability 
and processed data is in the range from one to four 8-hour shifts. 
Thus, the system operates 24 hours per day and seven days per week.

A second, identical farm of processors is used for data reprocessing. 
Using the same mechanisms as OPR, the reprocessing system is used to 
reprocess older data once newer code or improved constants become 
available.

\subsection{Computing System}
\label{sec:\secname:Computing System}
The \babar\ computing system includes the online system components
described above, plus the computing required for offline
reconstruction, simulation and analysis. This is implemented on several hundred
commercial Unix computers (currently running Solaris, Linux and OSF
operating systems), functionally
distributed, with tcp/ip-based networking. Fileservers with RAID disk arrays and
access to HPSS-based mass tape storage in STK robots are used to store and serve
the data. Network switching employs Cisco technology both online and in the SLAC
computing center. 
\par
The \babar\ software is mostly C++, designed with OOAD methodology. A commercial
database, Objectivity, is used for data and condition storage. Code distribution
employs \babar-developed release tools built upon CVS and the AFS distributed
file system. The LSF batch scheduling system is used at the SLAC site. Data
distribution to remote sites uses a combination of WAN, DLT7000 and
Redwood/Eagle tapes. Data acquisition and reconstruction is performed at SLAC.
Simulation and data analysis tasks are shared between SLAC and remote sites.
These remote sites include major centers in France, Italy, UK and USA.


\renewcommand{\secname}{Data}          
\section{Data Sample}
\label{sec:\secname} 

Results for this Conference are based on data collected up to  July 6, 2000, 
when the total integrated luminosity recorded by the experiment was
12.7\invfb. Of this sample, 11.3\invfb\ was taken ``on peak'', near the peak of the \FourS\ resonance, 
and 1.4\invfb\ was taken ``off peak'' at a center-of-mass energy 40\mev\ below the resonance. 
For the Osaka conference, the physics  analyses  are based on a sample of 8.9\invfb, of
which 7.7\invfb\ is on peak and 1.2\invfb\ off peak. An additional
1.1\invfb\ of on-peak data was used in the analysis for \stwob, for a total of 10.0\invfb.

The remaining 2.7\invfb\ of recorded data falls into two categories.
Data taken during the period between June and
October 1999 (1.3\invfb), 
when the DIRC system was incomplete (only 5 of the 12 bar
boxes were installed) and various subsystems were still optimizing their
operation, will be of limited use for future physics analyses. 
Some data (1.2\invfb) was excluded from the Osaka sample
because optimal alignment constants were unavailable when it 
was processed initially. 
This data will be available after reprocessing in August 2000.

\par



\subsection{Delivered and Recorded Integrated Luminosity}
\label{sec:\secname:Lumi}
The luminosity is monitored by \pep2\ and by \babar\ in the trigger and
prompt reconstruction.  
The L3 trigger calculates the deadtime,
which is typically less than two percent.  It also measures the luminosity
from Bhabha events.  The offline determination (described in
Section~\ref{sec:Analysis:Lumi}) uses dimuon and two-photon
events as well for cross-checks.  The DAQ efficiency is monitored closely
and is typically greater than 95\%.  Occasional scans of the resonance
are used to make sure that the experiment is running on the \FourS\ peak.

\subsection{Off-Resonance Running}
\label{sec:\secname:OffResonance}

By running below the \FourS\ resonance, \babar\ can study the non-
\BB\ contribution on-peak.  This is used to count the number of
observed \BB\ (see 
Section~\ref{sec:Analysis:BCounting}) and for background subtraction
in many analyses, especially inclusive measurements.  The amount of luminosity
devoted to off-peak running is of less importance for clean channels like
$\Bz \rightarrow J/\psi \KS$.  The uncertainty of a measurement with 
a ratio of signal plus background to signal $b$ depends~\cite{PhysicsBook} 
on the  fraction of continuum luminosity $c$ running as:
\begin{equation}
        \sqrt{\frac{c+b}{c(1-c)}} \ \ .
\end{equation}
For an analysis with $b=0.02$, collecting about 12\% of luminosity off-peak is optimal.  
This was the goal for the year 2000 run, which concentrates on
measuring \stwob.  As more luminosity becomes available, a somewhat
higher rate of continuum running may be chosen to improve the background
subtraction for modes like $\Bz \rightarrow \pip \pim$.

Operationally, the off-peak running took place about once a month.  This
minimized the amount of time spent tuning the machine after changing the
beam energy, while still allowing the analyses to track changes in machine backgrounds,
reconstruction code, and detector performance.  The center-of-mass energy
was reduced by lowering the HER energy by about 70\mev, giving a
center-of-mass energy 
of 10.540\gev. The LER  energy was left unchanged for 
simplicity's sake; the resulting decrease in boost was negligible for all analyses.

\subsection{Data Quality}
\label{sec:\secname:Quality}



Data quality is checked at each stage from the trigger to the
analysis skims. 
%
Gross hardware problems, such as power supply failures or
trips, are rare and are easily identified by the extensive slow control system.
A member of the shift crew is assigned to monitor
histograms, which are updated live with occupancy,
timing, event size, 
and other performance information from the L1 and L3 triggers, 
the detector subsystems, and special online event processing cross-checks.  
Automatic processes test additional live histograms
and report anomalies to the shifter or subsystem expert.  This insures that
problems are identified and fixed as soon as possible.
Several subsystems also employ automated processes
that perform tests on the data files to make long-term strip charts of
the kind of information listed above.


The reconstruction job contains many histograms for each subsystem
(including particle identification, beamspot and alignment monitoring, and 
results from some basic analyses).
A subset is inspected by subsystem experts for each run.  The most important
quantities are extracted from the histograms to make charts showing
the long-term behavior of the detector and reconstruction code, allowing detection of
changes that are 
too subtle or gradual to notice on a run-by-run basis.
The quality assessments from live data taking, online processing, 
and first-pass offline analysis are combined to define lists of
good runs for further analysis.  The yield per unit luminosity versus run 
for each
skim is inspected to insure that the data used in physics analyses
is of consistent quality within samples.
Less than one percent of data was declared unusable by this process.



\renewcommand{\secname}{Simul}          
\section{Simulation}
\label{sec:\secname}
The full detector simulation is structurally divided into three parts: event 
generation, the tracking of particles through the detector, and the detector 
response simulation.

\subsection{Event Generators}
\label{sec:\secname:Generators}
The simulation process begins by generating events using one of 
several possible event generators. The beam energies 
used in the simulation are thrown using a Gaussian distribution
of width 5.5 and 2.7\mev\ for the high and low energy beams 
respectively. A transverse momentum spread of 2.8 and 1.3\mevc,
obtained from beam emittances for the
HER and LER respectively, is also simulated. 
The position
of the interaction point is modeled
as a Gaussian distribution having widths of 125\mum, 4.2\mum, 
and 0.85\cm, in the $x$, $y$, and $z$ directions respectively.
The direction of the simulated beam axis is rotated with respect to the
\babar\ detector coordinate system
by $-18.8$\mrad\ in the $y$ direction, as observed
in data.

The main generator for \BB\ physics is called \evtgen, details
of which can be found in~\cite{ch4:evtgen}. This generator
provides a framework in which specific decay models can be 
implemented as modules. These modules can perform a variety of different
functions such as calculating amplitudes for the decay. 
\evtgen\ introduces mixing by generating decays of the
$\Upsilon(4S)$ to the proper mixture of \BzBzb,
$\bzb\bzb$, and $B^{0}B^{0}$ final states, with the
correct distributions in $\Delta t$.
\CP\ asymmetries are generated in modules that 
modify the generated lifetime distributions of the 
two $B$'s produced in the decay of the $\Upsilon(4S)$. Generic
models are available for simulating two-body decays to a pair
of scalar mesons, a scalar and a vector meson, a scalar and
a tensor meson, or a pair of vector mesons. 

A main decay file is used (DECAY.DEC) which 
provides a fairly complete table of the decays of particles 
below the $\Upsilon(4S)$ to exclusive final states. 
For generic decays of $B$ mesons, about 50\% are generated 
to exclusive states, 
while the other 50\% are produced using \jetset\ fragmentation model.
\jetset\ is also used to generate the \ccbar\ states and 
weakly decaying baryons. The original \jetset\ decay table was modified
in order to be in line with recent measurements.

\subsection{The GEANT3 Model}
\label{sec:\secname:BBSIM}

\bbsim\ is based on the 
detector description and simulation tool, \geant\ \cite{geant}, 
developed at CERN. 
\geant\ provides tools to construct the detector geometry; to step charged and
neutral particles through the detector; to simulate the full variety of 
interactions and decays that each particle species may undergo; to register 
Monte Carlo track hits (referred to as GHits\ in \babar\ terminology); and
to display the detector components, particle trajectories and track hits.

\bbsim\ is organized as a set of subsystem 
packages, each consisting of a standard set of routines which are called
at various stages of the event simulation. In the initialization
phase, the 
detailed subsystem geometries are built from parameters specified in an
ASCII geometry database.
The model includes the definition of 
the shapes, positions
and orientations of all the subdetector components, and of the materials
used in their construction.

Long-lived particles (``primaries'') produced in the event 
generation stage are stepped through the detector and allowed to interact or
decay. In \bbsim, hadronic interactions may be simulated using either 
\gheisha~\cite{gheisha} (default), \fluka~\cite{fluka}
or \gcalor~\cite{gcalor}.
Secondaries resulting from interactions or decays are tracked as well.
However, when recording the Geant3-level genealogy, 
only the kinematics of primaries and secondaries
resulting from decays are included.

Particle GHits are scored in the active detector components. The GHit
contains all the information needed to perform the subsequent detector 
response simulation, such as the position of the hit, 
the direction the track is traveling in, the energy lost, and the Monte
Carlo track number.All of subsystem GHits
are written to an output file, along with the generator- and \geant-level  
Monte Carlo event generation information. 

\subsection{The Detector Response Simulation}
\label{sec:\secname:SimApp}
The digitization of track hits is done as a separate step called \SimApp.
This uses the GHit information as input and outputs digitized data in
the same format as the real detector. At the end of this stage the Monte
Carlo data is processed by the same code that is run on the real
data. The code is again organized as a set of subsystem packages. These
packages contain routines to make the simulated data as realistic a
match as possible to data taken by the detector. As an example, the
drift chamber package takes account of the hit efficiencies measured
using real data. Several figures showing the quality of the agreement
between simulation and real data can be found in this paper. 

Another function of the code in the \SimApp\ packages is to add in
background hits. Rather than simulate the backgrounds in the detector, 
events from samples recorded with random triggers are overlayed (using 
the appropriate luminosity weighting factor) with Monte Carlo
simulated events. 

Simulation production takes place at SLAC and at remote sites. The
rate of production has been increasing with time and currently the level is
about three million events a month. Since background data is used in the 
simulation, and the backgrounds in the machine fluctuate, some care is 
needed to ensure that the Monte Carlo is produced with representative 
backgrounds. Each production block corresponds to about a month of real 
data and uses background files and detector efficiencies ``typical'' of 
that month. The background events are
shuffled to ensure that even small Monte Carlo data sets are generated
with a sample of background events spread out over the entire time period.


\renewcommand{\secname}{Tracking}          
\section{Tracking and Vertexing}
\label{sec:\secname}
\subsection{Track Finding and Kalman Filtering}
\label{sec:\secname:Kalman}
Charged track finding in \babar\ reconstruction starts with pattern  
recognition
in the DCH. A better event time is needed than that available  
from the L3
trigger, so the first step is to find an improved $t_0$ 
using three parameter fits
$(d_0,\ \phi_0,\ t_0)$ to four-hit track 
segments in the DCH superlayers.

The first pass at DCH track finding takes the track parameters and  
associated
lists of hits from the output of the L3 track finder and performs  
helix fits;
if a fit succeeds, that track is retained. A search for 
additional hits in the DCH that
may belong on these tracks is made, and the event time is  
further refined
using the timing information of just the hits assigned to tracks. Two 
subsequent track finders are run on the remaining DCH hits. Both  
track finders
work on superlayer segments constructed out of hits not used on previous 
tracks, and are designed to find tracks at lower \pt\ than found  
by the L3
trackfinder, to find tracks going through fewer than the full ten  
superlayers,
and to find tracks without bias towards the interaction point.  In  
this way
more sophisticated track finders run in a progressively cleaner tracking 
environment with a constantly improving measure of the event time.

At the end of this process, the found tracks are refit using a  
Kalman filter
fitter.~\cite{dchtrack}
This fitter takes into account the detailed distribution of 
material in the detector and the full map of the non-uniformities  
in the detector magnetic field. The tracks are then projected into the SVT, and  
silicon strip hits are
considered for addition to the tracks if they are consistent with  
the expected
error from extrapolating through the intervening material and field.
Various possible combinations (branches) of SVT hits are considered, and
the branch contributing the least to $\chi^2$, while meeting pattern 
requirements designed to limit the number of silicon wafers providing no 
measurement, is retained. When all intersected wafers have been  
inspected for
hits and the best branch is selected, a full Kalman fit is  
performed on the
ensemble of DCH and SVT hits.

Then a search 
for tracks is performed using the
remaining unused silicon strips, as  
described
in the next section. 
At the end of the SVT-only track finding, a search of the track list
is made for tracks made up of only SVT hits which match closely  
at the
support tube  
to other
tracks made up of only DCH hits, since the
support tube is a localized source of scattering between the two devices.
 If any  
candidates are
found, a common fit to both sets of hits is attempted. If this  
succeeds, only
the combined track is retained; otherwise, both stand-alone tracks are  
kept on the
track list. At the end of pattern recognition and fitting, all  
tracks are
checked to see if they lie in momentum ranges where the particle  
hypothesis
might effect the result of the fit, and if so, each track is fit  
with up to
five stable-particle mass hypotheses.

\subsection{Tracking Efficiency}
\label{sec:\secname:TrkEfficiency}

The absolute level of track reconstruction efficiency has been determined by studies using
control samples from data. Changes in the DCH operating point and the
reconstruction code during data taking are also followed in this way. The polar and
azimuthal angle dependence of the track finding efficiency is modeled through detailed
tuning of the Monte Carlo simulation. The overall level 
of track reconstruction efficiency and its dependence on angle or \pt\
is checked using:
\begin{itemize} 
\item
Bhabha scattering
events at low multiplicity and high \pt; events are selected based on calorimeter information and the efficiency for finding the \epem\ tracks determined;
\item
one-versus-three tau-pair events at moderate
multiplicity and \pt\ ranging up to about 2\gevc; events are selected without requiring
the third track on the 3-prong side, allowing a determination of the probability
for reconstructing this track with a variety of track selection requirements;
\item
events containing
a $\Dz\to\Km\pip\pip\pim$ decays where the $\Dz$ is reconstructed 
from the decay $D^{*+}\to\Dz\pip$; the relative tracking efficiency for tight or loose
requirements on the DCH portion of the track can be determined, as well as
absolute efficiencies using the ratio of $\Dz\to\Km\pip\pip\pim$ to $\Km\pip$;
\item
multiplicity distribution of clean multihadron event selections;
\item
comparisons between standalone tracking in the SVT and DCH.
\end{itemize}
We assign a conservative 2.5\% systematic error per track to the absolute
track finding efficiency
above 200\mevc\ in \pt\ in computing all branching fractions reported at this conference,
and treat this as fully correlated for all studies that require a DCH portion
of the track. This represents the level of consistency among track finding efficiency
studies done so far, but in no way reflects the final systematic error that we
expect to achieve with the tracking system. 

\subsection{Slow Pion Reconstruction}
\label{sec:\secname:SlowPi}

%
%


   The tracking system in \babar\ was designed to be highly efficient
even for very low momentum tracks, which are important for 
reconstructing decays such
as $D^{*+}\rightarrow D^0 \pi^+$ for which the $\pi^+$ has a very soft
momentum spectrum.
Tracks with transverse momentum (\pt) below about 120\mevc\ can not
be found with the DCH pattern recognition because they
curl up and do not penetrate deep into the DCH, which
has an inner radius is 24\cm.
The five-layer SVT was designed to be a stand-alone tracking device 
covering specifically the entire \pt\ range of 50--120\mevc.
The lower \pt\ limit of 50\mevc\ is the threshold below which pions
have a large probability of stopping in the material of the 
beam pipe.

The SVT stand-alone pattern recognition is performed 
by two independent algorithms after
SVT hits have been associated with, and claimed by, stand-alone
DCH tracks.
The first algorithm builds tracks from space points, which are
intersections of $\phi$ and $z$ strips.
The second algorithm forms circle trajectories with the
$\phi$ hits and then adds $z$ hits to the circles to form
helices.
The first algorithm is efficient over a wide range of 
impact parameter and $z_0$ values
but is susceptible to inefficiencies
in events with an extremely large number of hits
and from silicon modules with non-functional sides 
($\phi$ or $z$).
The second algorithm is less sensitive to extreme combinatorics
and module defects and is used to find tracks missed
by the first algorithm.

\begin{figure}[ht]
\setlength{\epsfxsize}{0.95\hsize}
\centerline{\epsfbox{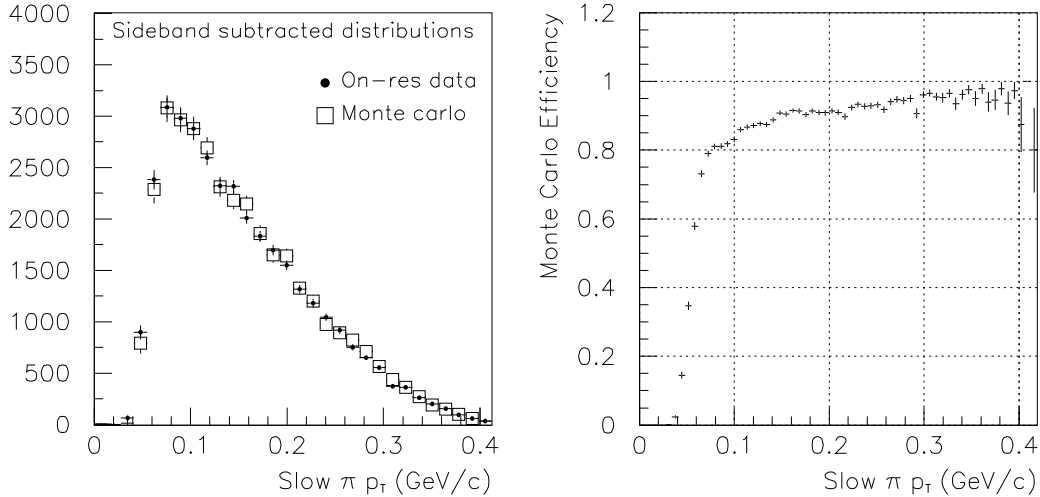}}
\begin{center}
\parbox{5.5in}
{\caption
{\label{fig:slow-pi-fig}
The figure on the left shows the slow pion
\pt\ spectrum, after subtraction of combinatorial background using the 
$m(D^0)$ sidebands, for on-resonance data (points) and Monte Carlo (boxes).
The plot on the right shows the tracking efficiency as a function of
\pt\ derived from the Monte Carlo.
}}
\end{center}
\end{figure}

Figure~\ref{fig:slow-pi-fig} shows the slow pion
\pt\ spectra from the decay $D^{*+}\rightarrow D^0 \pi^+$,
after subtraction of combinatorial background using
the $m(D^0)$ sidebands,
for
on-resonance data and a mixture of \bbbar\ and \ccbar\ Monte Carlo
which represents the on-resonance conditions.
The shape of the \pt\ spectra agree quite well, which
gives us some confidence in our Monte Carlo simulation.
Figure~\ref{fig:slow-pi-fig} also shows the tracking efficiency as 
a function of \pt\ as derived from the Monte Carlo.
The efficiency turns on at a \pt\ of about 50\mevc\ and crosses
80\% at about 70\mevc.
We have tried to measure the relative tracking efficiency at low
momentum from the data alone using
slow $\pi$ helicity distributions, binned in 
$p^*(D^{*+})$.
%
Currently, this measurement is statistically limited, but will
eventually yield a tracking efficiency measurement at low momentum
from the data without Monte Carlo dependence.
We assign an
uncorrelated 2.5\% systematic error to the track finding efficiency for those tracks
that rely on the SVT only. 

%
%

\subsection{Vertex and Kinematic Fitting}
\label{sec:\secname:Vertex}
\def\D     {\ensuremath{D}}
\def\Dstarp    {\ensuremath{D^{*+}}}
\def\Dstarm    {\ensuremath{D^{*-}}}

Vertex and kinematic fitting is widely used to improve four-momenta and position
measurements, as well as to measure the time difference between 
decaying \B\ hadrons in the $\FourS \to \B\Bbar$ decay. Combined
with tracking and particle identification, vertex and kinematic fitting provides the
basis for event identification.

The physics properties of a decay are used
to apply constraints which translate to better mass and position resolutions and
larger signal-to-background ratios. For example, in the case of
\bpsiks, the position measurement of the \Bz\ can be improved using the
fact that the line-of-flight of the \KS\ intersects the \jpsi\ vertex.
The energy resolution of the \Bz\ can also be improved by applying 
a mass constraint to the \jpsi. 
Several constraints have been
implemented utilizing the Lagrange Multiplier
mechanism \cite{VertexRef}: common decay vertex, mass, energy, momentum,
beam energy (with and without smearing), beam spot position and
line-of-flight. In all cases constraints are applied in three dimensions.


One of the fundamental principles in the design was to deal in a simple way
with complex decay chains. Virtual {\em composite} particles and
their error matrices are built from the original particles.
The composite particle then replaces
the daughters in subsequent fits and analysis. To account for
intermediate resonances and neutral particles, vertices can be shared by different virtual states. 
The three-momentum of the virtual
particle is fit directly, rather than computed from the updated
daughters, improving speed and numerical accuracy.

Non linearities in the fits require the application of
an iterative procedure. Simple fits involving only vertex
constraints (except long-lived particles, called $\ensuremath{V^0}$'s)
are, however, accurate enough with a single iteration. This has an important
impact on the amount of time consumed in vertex fitting. The other fits 
involving kinematic constraints and $\ensuremath{V^0}$'s require, in
general, more than one iteration. The criteria we use for convergence
is that the change in $\chi^2$ between two successive
iterations is less than 1\%, with a maximum of six iterations.
In all cases the initial location of the vertex is estimated by
solving analytically (with an iterative procedure) the point of
closest approach of the two tracks, using a second-order approximation
at each point.

Track and vertex parameter errors depend heavily on our understanding internal and global alignment, 
detector material, and magnetic field variation. A set of control samples 
($\gaga \to 4\pi$, $\tau \to 3$-prongs, $\Dz \to \Km \pip \pim \pip$, $\Bub \to\jpsi \Kp$) has been
used to estimate from data the vertex position and mass
resolutions and systematic uncertainties in their determination.
The pull distributions of these control samples show that tracking and
vertex errors are estimated correctly at the
10--20\% level and that biases in determining
vertex positions due to alignment problems are at the level of a few microns.

\subsubsection{Beam spot and primary vertex determination}

In order to accommodate the movements of the \pep2\ beam with 
respect to the \babar\ detector,
the position of the collision point at \pep2\ is determined
from track-based methods. The
apparent size from two-prong events is about 150\mum\ in $x$, 50\mum\
in $y$ and 1\cm\ in $z$. The apparent size in the $y$ direction is totally
dominated by the track resolution. A better estimate of $\sigma_y$
can be obtained from the knowledge of the
luminosity, the beam currents and the size in $x$, providing a value
of about 4\mum, varying within 10\% on a time scale of hours.

For many physics analyses and
reconstruction tasks, it is necessary to have a more accurate estimate
of the collision point in $x$ and $z$.
For instance, reconstruction of $\gamma$ and
$\pi^0$ candidates uses, in many cases, the primary vertex as the point of origin in
reconstructing vectors. The primary vertex\footnote{
For light quark events, the reconstructed vertex is a good estimate of the true primary
vertex. For \BB\ events, where both \B\ mesons travel along the $z$ axis
in the laboratory frame, the primary vertex provides an average \B\ decay
position.
} 
is estimated on an event-by-event basis from a
vertex fit which uses charged tracks with an impact parameter (with respect to the beam spot
position) less than 1\mm\ in the transverse plane. Tracks with high
$\chi^2$ contribution to the vertex fit are removed until an overall 
$\chi^2$ probability greater than 1\% is obtained. The
resolution is about 70\mum\ in $x$, $y$ and $z$ for hadronic events.

\subsubsection{Reconstruction of the tagging-\B\ vertex}

In measurements of the \B\ lifetime, mixing rate and time-dependent
\CP-violating asymmetries, we fully reconstruct or partially
reconstruct only one \B\ and then determine the distance between the
two \B\ decays. Determination of the fully-reconstructed \B\ decay
vertex makes use
of all the final-state tracks of the \B\ decay chain.
The other \B\ vertex (tagging-\B\ vertex) is reconstructed
with the remaining charged tracks in the event. To retain high
efficiency this is done with inclusive techniques.
Complications arise from the fact that one has to deal with
secondary tracks from short- and long-lived particles. To deal
with these problems and minimize their impact,
the three-momentum of the 
tagging-\B\ (called a pseudo-track) and associated error matrix are 
derived from the
three-momentum, decay
vertex and error matrix of the fully reconstructed \B\ candidate, 
and from a knowledge of the average
position of the interaction point and the $\Upsilon(4S)$
four-momentum. In the case of partial \B\ reconstruction (where the
three-momentum is not known) the constraint from the
\B\ pseudo-track cannot be used.
This pseudo-track is fit to a common vertex with the remaining tracks
in the event. Reconstructed $V^0$ candidates are also used in this
procedure, reducing biases from long-lived decay particles.
Tracks with contributions to the $\chi^2$ greater than
six are removed from the fit. The procedure is iterated until there
are no tracks contributing more than six units to the $\chi^2$
or all tracks are removed. 

The $\Delta z$ resolution function obtained with this technique can be 
accurately described by the
sum of two Gaussian distributions with different means and
widths. Typically, 80\% of the events are in the core Gaussian, which
has a width $\approx 100$\mum. The remaining events reside in
the tail Gaussian which has a typical width $\approx 350$\mum.
The bias due to secondary tracks is $\approx -20$\mum\ for the core 
Gaussian and 
$\approx -80$\mum\ for the tail Gaussian. An inclusive
$\Dstarm \to \Dz \pim$ control sample (selected from
continuum and off-resonance running) is used to check the $\Delta z$
reconstruction and resolution for data.


\renewcommand{\secname}{PID}          
\section{Particle Identification}
\label{sec:\secname}


\subsection{Charged Particle Identification}
\label{sec:\secname:Charged}
\subsubsection{Electrons}

\begin{figure}[!htb]
\begin{center}
\includegraphics[height=2.5in]{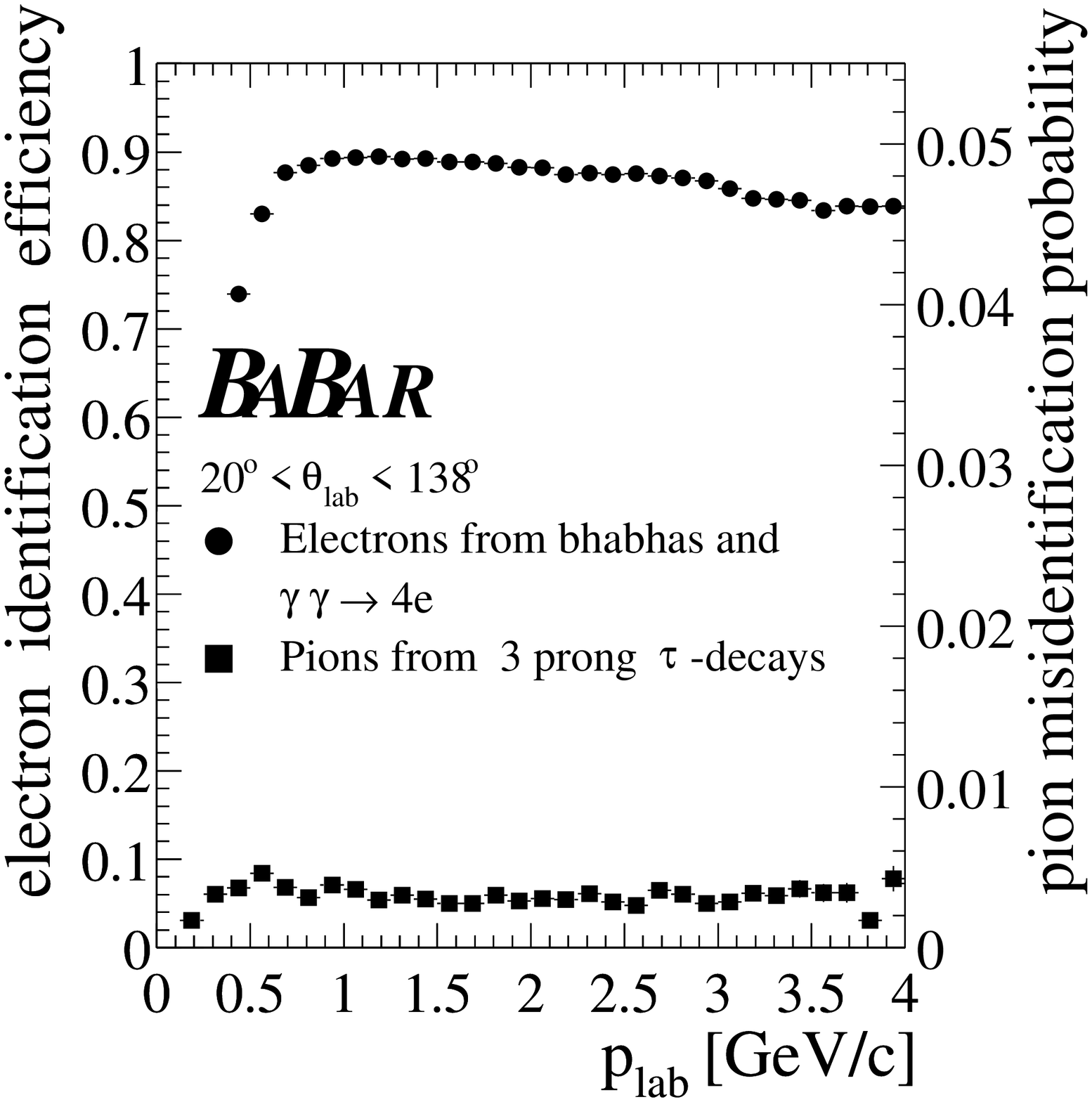}
\hskip2truecm
\includegraphics[height=2.5in]{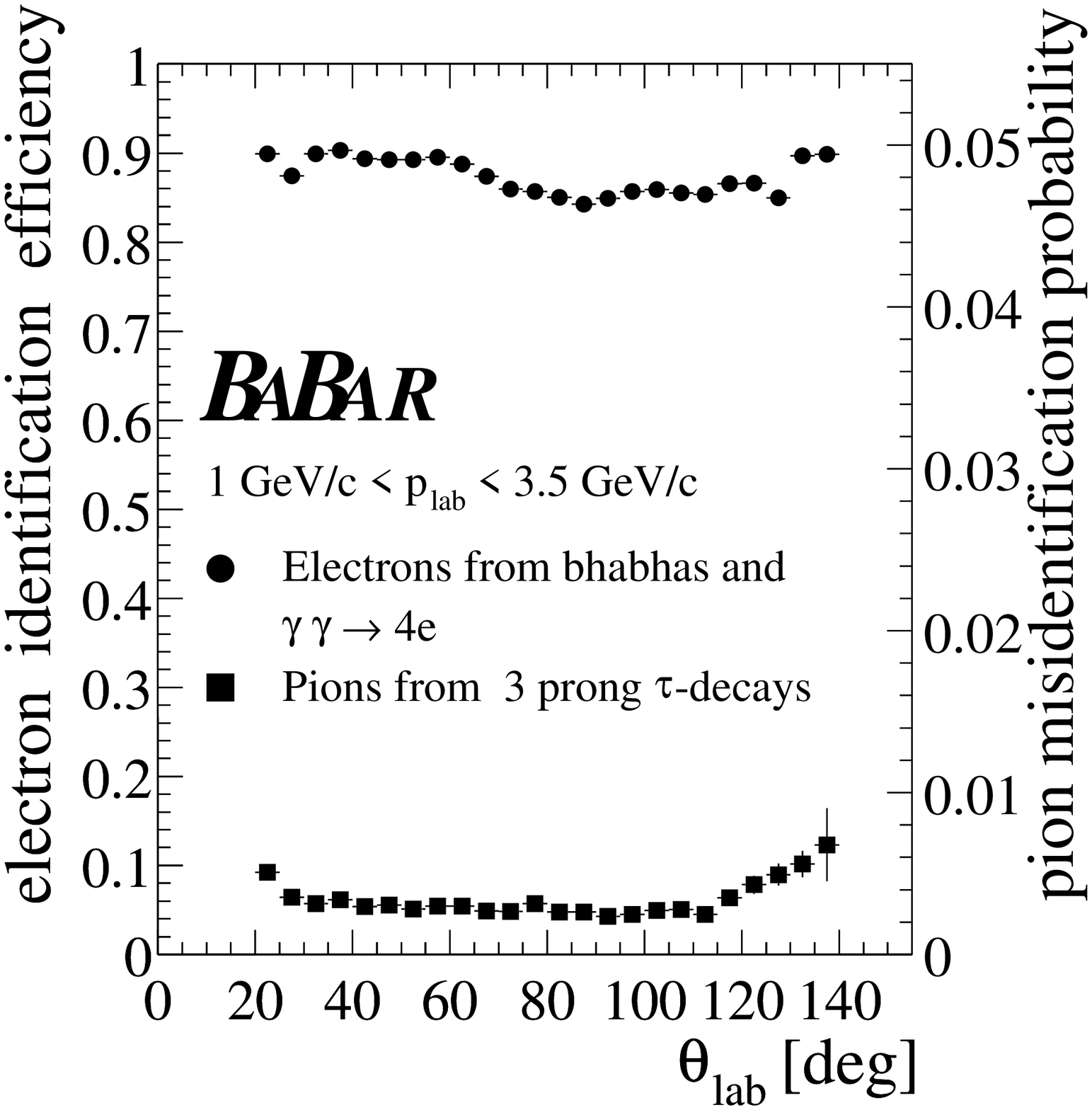}
\caption{Electron   identification    efficiency  and   pion
misidentification  for the very tight selection  criteria as a
function of momentum (left) and polar angle (right) 
(note the different scales for electrons and pions). }
\label{fig:electroneff}
\end{center}
\end{figure}

Electrons are identified with the EMC, the
DCH and the DIRC systems, using quantities
such as 
the ratio of the deposited energy over the momentum, and variables
describing  the spatial
shower development. The lateral shower development is measured with a
method introduced  by ARGUS~\cite{drescher} and  expanded into various
moments   containing   information   about  the   shower's   azimuthal
distribution    with    respect     to    the    particle's    initial
momentum~\cite{sinkus}.   A cut is  applied on  the energy  loss 
using a truncated mean of 40 samples  (maximum) recorded in the
DCH. The measured Cherenkov angle in the DIRC is required to 
be within
$3\sigma$  of the  expected value  for the  electron  hypothesis.  The
efficiency of  electron identification  is studied by  using electrons
obtained   from  radiative   Bhabhas   and  events   of  the   process
$\gamma\gamma\rightarrow  e^+e^-$.  The  misidentification probability
for pions is  measured with 3-prong $\tau$ decays.   The {\it tight\/}
selector has an  average efficiency of $94.8\%$ in  the momentum range
$0.5\gevc < p < 2 \gevc$, with a pion misidentification probability of
roughly $1.2\%$.  The {\it very tight\/} selector has an efficiency of
88.1\%  with   an  average  pion  misidentification   of  0.3\%.   The
dependence of the  efficiency on momentum and polar  angle is shown in
Figure~\ref{fig:electroneff}.

%
 

\subsubsection{Muons}

Charged tracks reconstructed in the DCH are extrapolated 
to the flux return iron using a detailed map of 
the non-uniform magnetic field and accounting for
the expected average energy loss.
The predicted average position of the intersections with the active 
detector planes is computed, 
including the uncertainty due to multiple scattering.
All hits found in each readout view within a specified maximum distance 
from the predicted
intersection are associated with the charged track.
We require, for tracks within the acceptance of the EMC, 
the energy deposition to be consistent with that 
expected of a minumum ionizing particle.
A signal in at least two layers of the IFR is also required. A cut is also made
on the difference between measured and predicted 
(based on the muon hypothesis)
total number of interaction lengths traversed
in all subdetectors. We expect the average number of signal strips per layer
to be larger for pions produced in an hadronic interaction than for muons.
The average value and the r.m.s. of the strip pattern for the different hit layers
provides $\mu/\pi$ discriminating power. Finally, we reject tracks with a large 
track-match $\chi^2$ or a large $\chi^2$ for a polynomial fit to the IFR cluster.  

\begin{figure}[!htb]
\begin{center}
\mbox{\epsfig{file=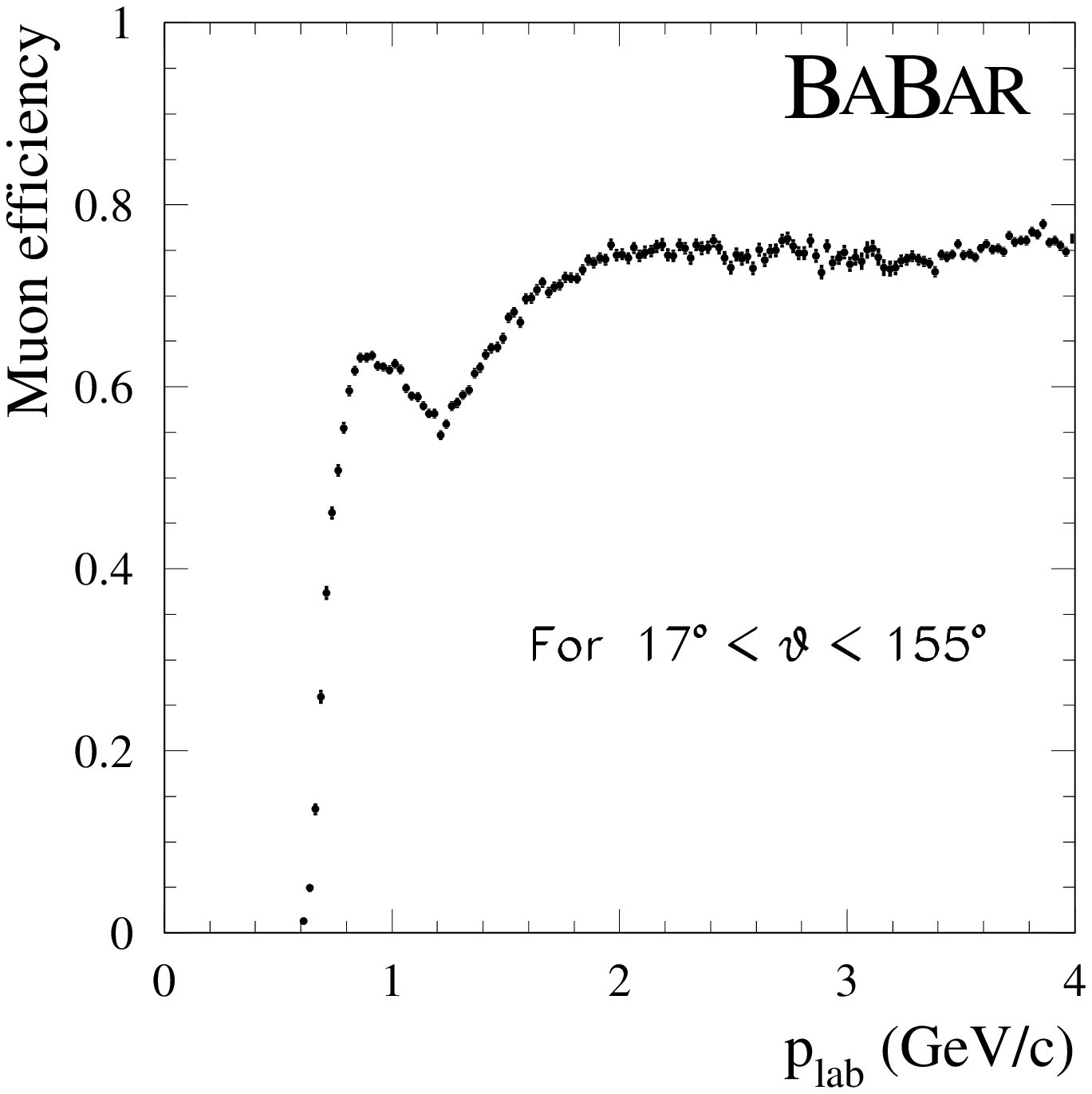,width=0.45\textwidth}}
\mbox{\epsfig{file=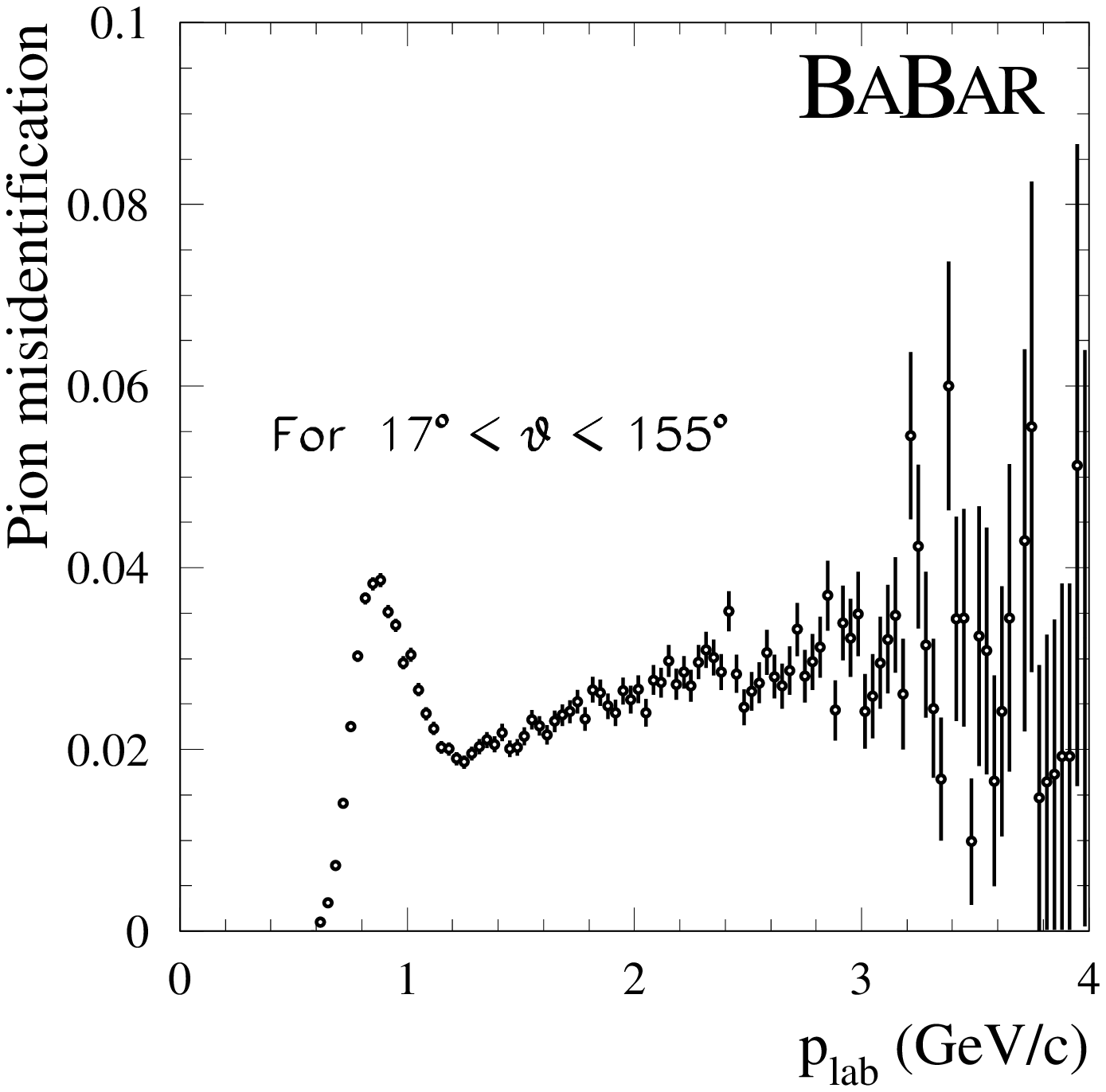,width=0.45\textwidth}}
\end{center}
\caption{
Muon identification efficiencies (left) and pion misidentification rates (right) for the tight muon selector, 
showing average efficiencies of 75\% and misidentification rates of 2.5\% for the momentum range 
1.5 to 3\gevc.
}
\label{fig:muon}
\end{figure}

The performance of the muon selectors has been tested using control samples obtained 
from data. 
The processes considered are $\mu\mu e e$ and $\mu\mu\gamma$ 
for muons, and $\tau\rightarrow 3$-prongs and $K_S \rightarrow \pi^+\pi^-$
for pions.
The selection of the control samples is based on kinematic variables and
particle identification.
No muon requirements are applied to the tracks that are used as a control sample, in order 
to avoid any bias in the determination of the particle identification performance. 
Different muon selection criteria with different levels of purity are defined for analysis 
specific applications. 
In the range $1<p<3$\gevc\ the average efficiency is about $77\%$ for 
muons with a $2.5\%$ pion fake rate (see  
Figure~\ref{fig:muon}).


\subsubsection{Charged hadrons}

Kaons are selected using the information from the SVT, DCH, 
and DIRC systems.
Likelihoods are calculated from the \dedx\ measurements in the
SVT and DCH, and from the angle and number of photons found
in the Cherenkov ring in the DIRC. The individual likelihoods from the three detectors 
are then multiplied together and a cut on the ratio of likelihoods (K vs $\pi$, K vs
p) is applied:
\begin{equation}
L_{\pi} \cdot r < L_{K} \,\,\,\mbox{and}\,\,\, L_{p}\cdot r < L_{K}
\end{equation}
Various kaon selections have been implemented, achieving different levels of
efficiency and purity. Each detector output is used only in a specific
interval in momentum. For example, for a tight selection, the ranges used for
each subdetector are 0.025--0.7\gevc\ for SVT, 0.090--0.7\gevc\ for DCH, and $>0.6$\gevc\ for the DIRC.
The minimum requirements for a track to be considered in the
individual subdetectors are 2 and 10 \dedx\ samples for the SVT and DCH respectively, and track extrapolation
within the DIRC acceptance (number of expected photons greater than zero).
Finally, the value of likelihood ratio $r$ was dependent on the momentum of the
tracks to be identified. In the tight mode, the intervals and ratio requirements were:
\begin{center}
\begin{tabular}{cl}
0.5 - 0.7\gevc & $r = 15$ \\
0.7 - 2.7\gevc & $r = 1$  \\
$>2.7$\gevc    & $r = 80$ \\
\end{tabular}
\end{center}

\begin{figure}[!htb]
\begin{center}
\epsfxsize=3in   
\epsffile{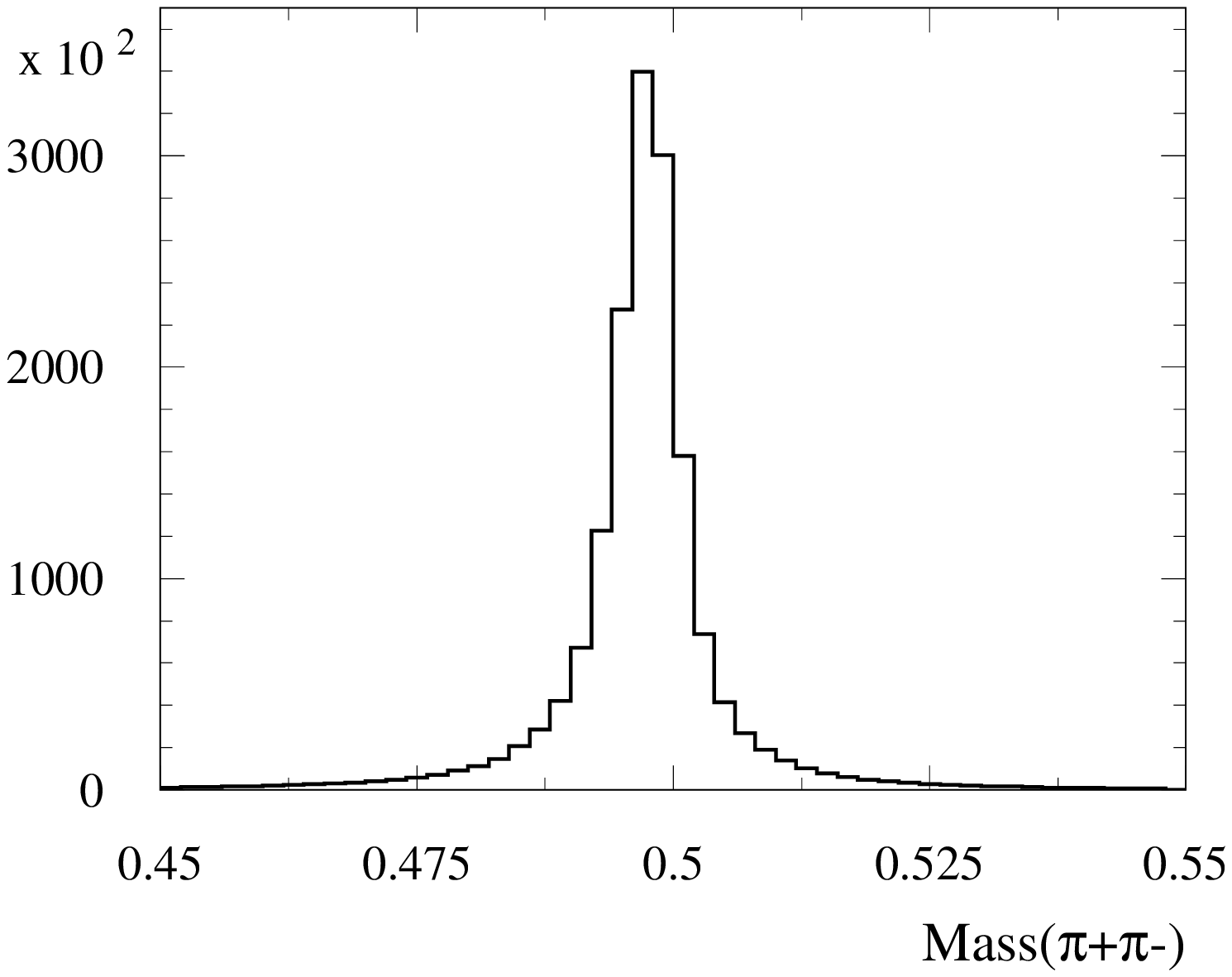}
\epsfxsize=3in   
\epsffile{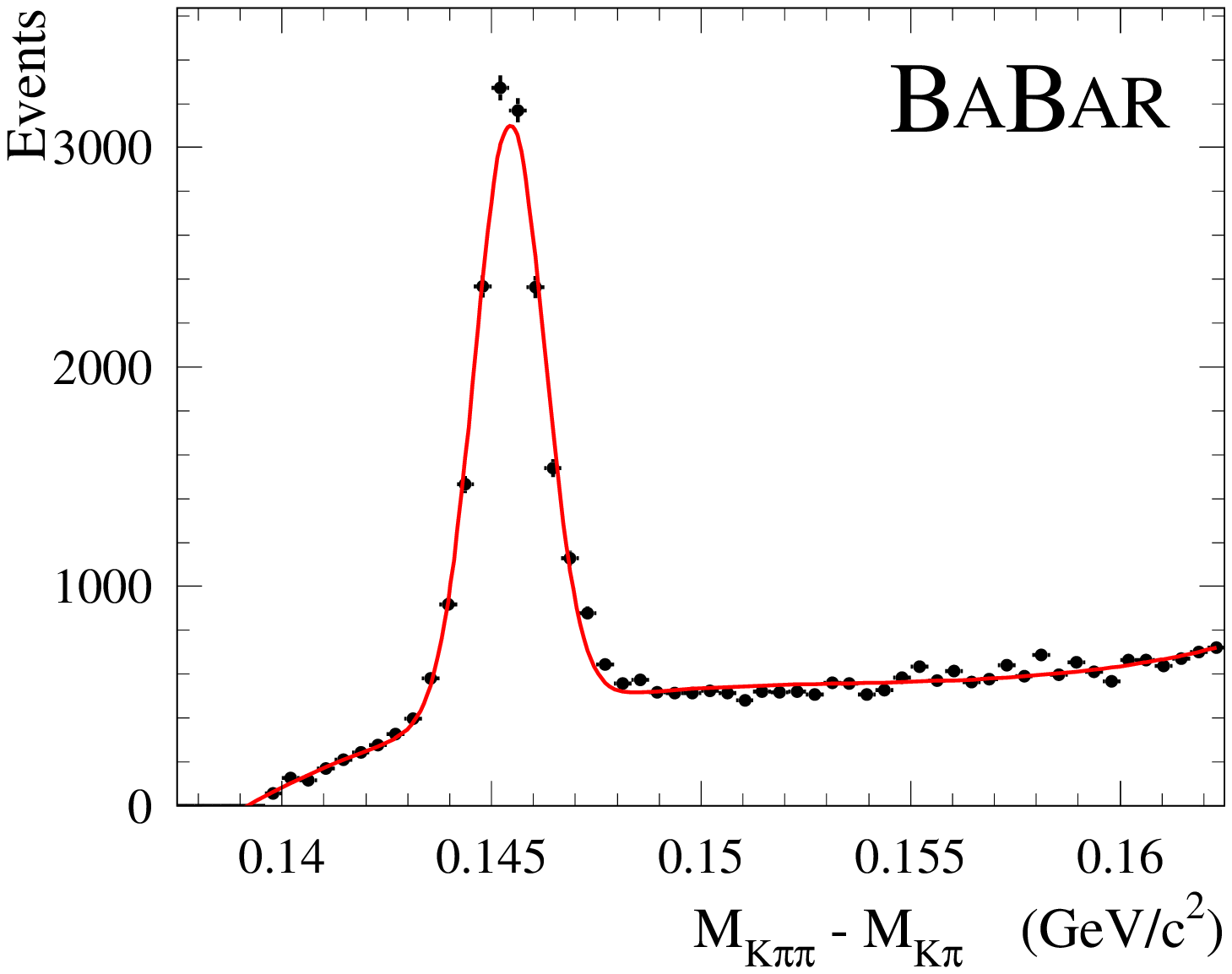}
\end{center}
\caption{Mass distribution (left) of $K_S^0$ candidates used to select a pion
control sample.
Distribution (right) of the difference of the $D^{*}$ and $D^0$
candidate masses, used to select control samples of kaons and pions.
} 
\label{Kshort}
\end{figure}

The performance of the various selections has been measured with data using
control samples of kaons and pions from the kinematically identified
decays: $K^0_S\rightarrow \pi^+\pi^-$ and $D^{*+}\rightarrow
D^0(\rightarrow K^-\pi^+)\pi^+$. 
To select the \KS\ sample, very tight cuts are used (e.g. on the angle between
the $K^0_S$ candidate vertex direction and the vertex momentum direction,
on the decay distance from the interaction point, on the reconstructed
$K^0_S$ candidate mass). The plot of the invariant mass of the
$\pi^+\pi^-$ pairs is shown in Figure~\ref{Kshort}(left); the purity of the
resulting pion sample is $>99\%$.   
The 
$D^{*+}$ sample, obtained 
with a tight cut on the difference between
the $K\pi\pi$ and $K\pi$
masses ($0.1445 <\Delta M< 0.1465$\mevcc)
shown in
Figure~\ref{Kshort}(right),
has a combinatorial background of around 
$\sim13\%$ (for a kaon sample purity of $\sim90\%$, with the purity 
higher at higher
momenta). In determining particle identification performance, a correction
for the
background contribution has been made in each   
momentum bin using sideband subtraction.

\begin{figure}[!htb]
\epsfxsize=3.0in   
\centerline{\epsffile{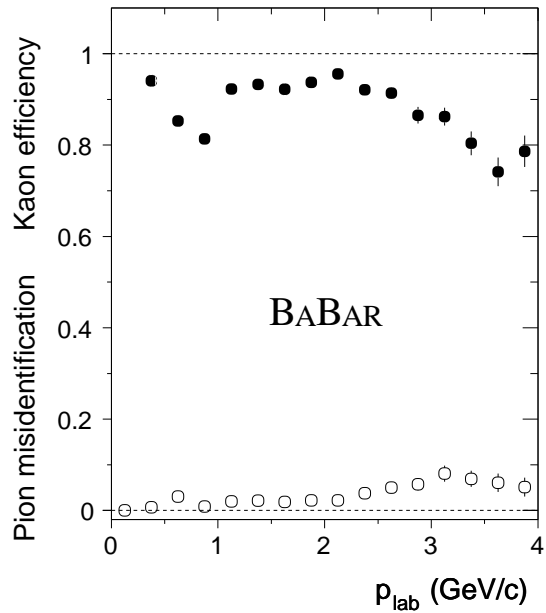}} 
\caption{Kaon efficiency (filled points) and pion misidentification (open points) for 
a tight kaon selection, measured with a $D^{\star}$ decay control sample.} 
\label{EffMis}
\end{figure}

The measured kaon efficiency and pion
misidentification rates are shown in Figure~\ref{EffMis} for the tight
selection based the $D^{*+}$ sample. The efficiency is found to be 90\%, while
misidentification rate is 2.5\%, averaged over the full momentum range.
Note that these results are highly dependent on the momentum spectrum of the kaons and
pions in the particular samples used in an analysis. 

\subsection{Neutral Particle Identification}
\label{sec:\secname:Neutral}
\subsubsection{Photons}
Photons are detected in the EMC where 
adjacent crystals with energy deposits exceeding 1\mev\ are grouped in 
clusters. Channels marked as noisy by online or offline monitoring are 
excluded. 
Clusters with more than one local energy maximum are then 
split into ``bumps'' and the energy of each crystal is partially assigned 
to each bump by a simultaneous iterative adjustment of the centers 
and energies of the bumps, asuming electromagnetic shower shapes. 
In the next stage all charged tracks reconstructed in the tracking 
volume are extrapolated to the EMC entrance and a track--bump matching 
probability is calculated for each pair. All bumps with a small matching 
probability are treated as photon candidates. A small number of these candidates, 
where the bump shape is incompatible with that expected of an electromagnetic shower, 
are rejected. 
In the majority of physics analyses, photon candidates lying 
anywhere within the instrumented EMC volume are used if their energy is above 
30\mev.

\subsubsection{\piz\ and $\eta$ Reconstruction}
\begin{figure}[!htb]
\begin{center}
\begin{tabular}{lr}
\mbox{\epsfxsize=7.2cm\epsffile{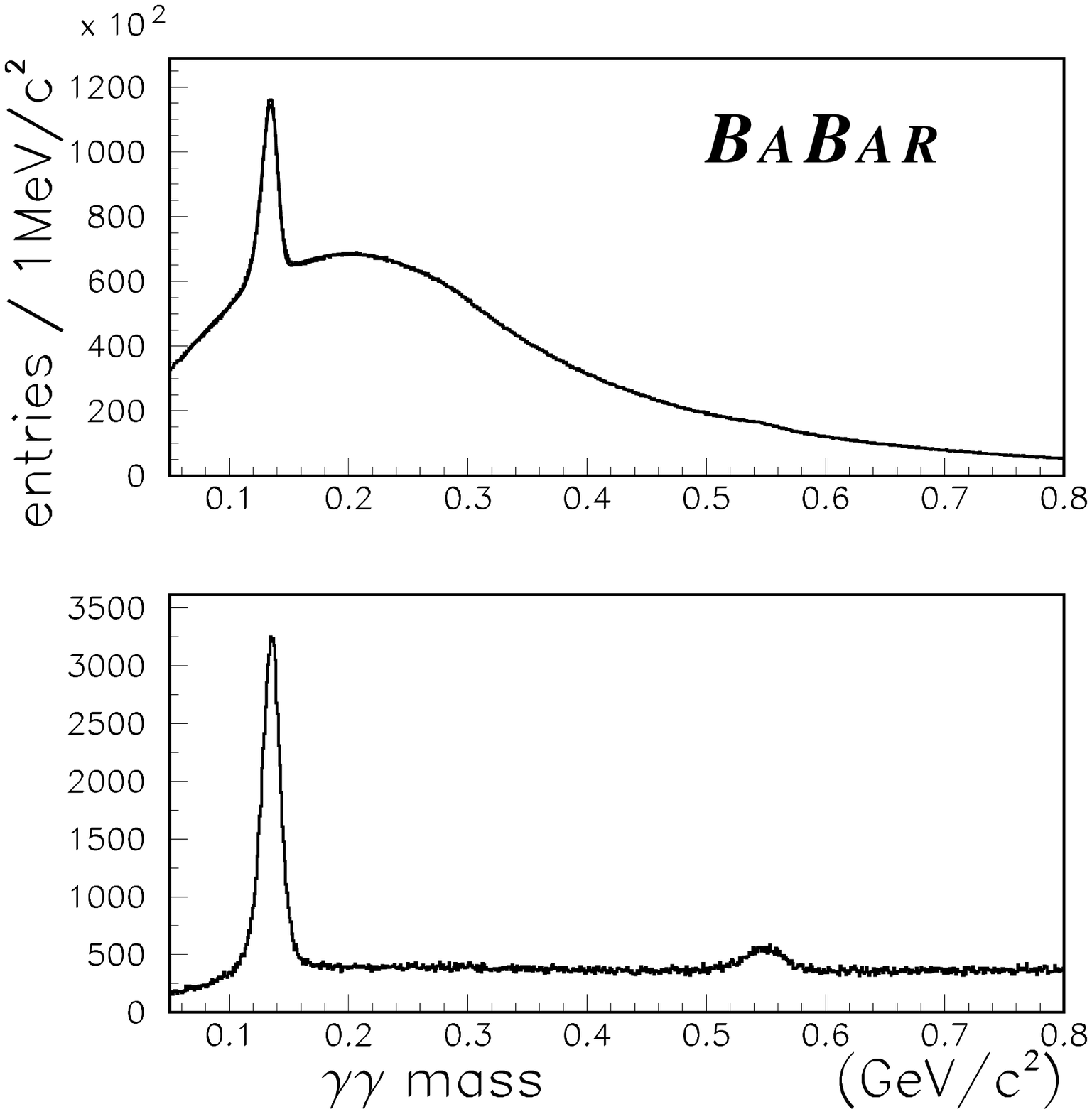}} &
\mbox{\epsfxsize=7.2cm\epsffile{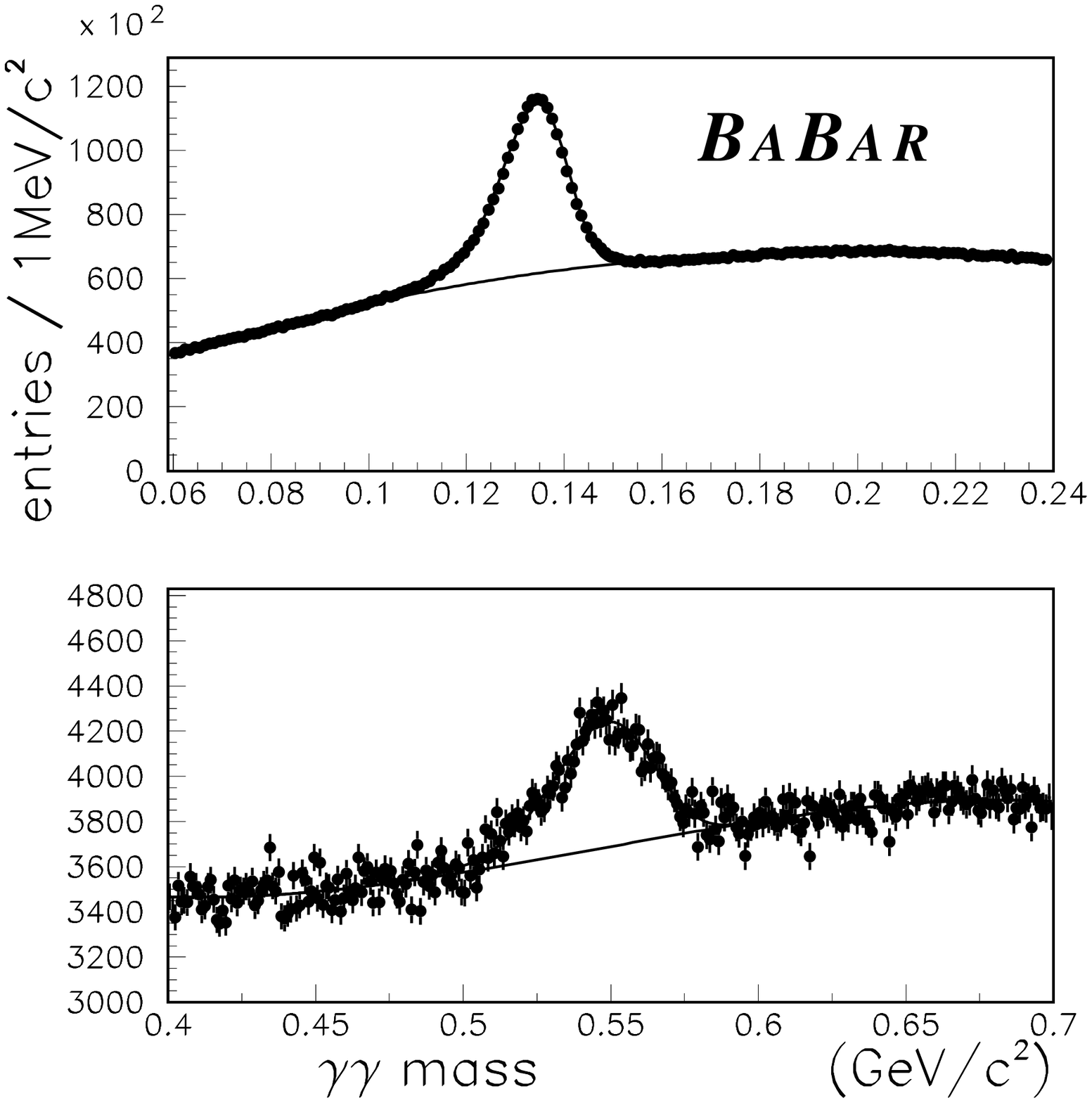}} \\
\end{tabular}
\end{center}
\caption{Invariant mass of $\gamma\gamma$ pairs from hadronic events. Upper pair of plots are for 
$\rm{E_{\gamma}} >$30\mev\ and $\rm{E_{\gamma\gamma}} >$300\mev, with enlargement of the \piz\ mass region on the right; 
lower pair of plots are for 
$\rm{E_{\gamma}} >$100\mev\ and $\rm{E_{\gamma\gamma}} >$1\gev, with enlargement of the $\eta$ mass region on the right.}
\label{fig:ggmass}
\end{figure}

Neutral pion and $\eta$ candidates are formed from pairs of photon candidates 
assumed to originate from the interaction point. 
The invariant mass spectrum of all such pairs is shown 
in Figure~\ref{fig:ggmass} for 
($\rm{E_{\gamma}} >$30\mev, $\rm{E_{\gamma\gamma}} >$300\mev) and 
($\rm{E_{\gamma}} >$100\mev, $\rm{E_{\gamma\gamma}} >$1\gev), where clear 
\piz\ and $\eta$ peaks can be seen. 
The \piz\ mass resolution in multihadron events is 6.9\mevcc, while in 
low occupancy $\tau\tau$ events it is 6.5\mevcc\ for \piz\ energies 
below about 1\gevc. Improved resolution is 
also observed in hadronic events where only isolated photons are considered. 
The $\eta$ mass resolution is 16\mevcc. The \piz\ and $\eta$ 
mass peaks from inclusive multihadron events are also shown in 
Figure~\ref{fig:ggmass}. The signal-to-background ratio for the $\eta$ 
can be improved by vetoing photons participating in candidates in the 
\piz\ signal region. Combinatorial background 
in hadronic events is supressed by a factor of 5 by this veto requirement, with a signal loss of only 25\%. 

The detector segmentation and achieved spatial resolution allow the 
reconstruction of \piz's with photon separation in the electromagnetic calorimeter as small as 5\cm, although with some
deterioration of the resolution. The small fraction of high 
energy \piz\ whose photons cannot be resolved into individual photon bumps
($\approx$10\% in the 4--6\gev\ 
region) can be separated from single photons using the cluster shape 
(the second moment of the energy distribution around the cluster centroid).

Large samples of inclusive and exclusive data have been used for simulation comparison
studies. The \piz\ mass value, resolution, and 
efficiency, and their (weak) energy dependence were found to be in very 
good agreement. Furthermore, the \piz\ mass, width, and yield, 
are monitored in real time during data collection and reconstruction. 
Limits on the possible discrepancies between the data and the
simulation for the photon energy 
scale, resolution, and reconstruction efficiency, have been 
conservatively set to be 0.75\%, 1.5\% and 2.5\% respectively.

\subsubsection{Neutral Hadrons}
\label{sec:\secname:NeutralHadrons}
\begin{itemize}
\item
\KS\ $\rightarrow$ \pipi



To reconstruct $K_S^0$ candidates, we consider any pair of oppositely-charged 
tracks combined in a common vertex.
No cut is made on the probability of $\chi^2$, but the vertex fit must converge.
The recontructed momentum of the $K_S^0$ candidate is required to be aligned 
with the direction between the interaction point and the reconstructed vertex.
Finally, a minimum transverse momentum of the daughters with 
respect to the flight direction is required in order to eliminate combinatorial and $\Lambda$ 
contamination.
For Figure~\ref{pid:neutralhad:ksmass}, a tighter selection 
is applied on a small data sample; the $\chi^2$ probability is
required to be 
greater than 0.01, the angle between the direction of flight and the direction 
from the interaction point to the reconstructed vertex is required to be
greater than 45\mrad,
the transverse momentum of the daughters with respect to the $K_S^0$ 
flight direction is required to be  
between 0.11 and 0.22 \gevc.

\begin{figure}[!htb]
\begin{center}
\includegraphics[height=2.5in]{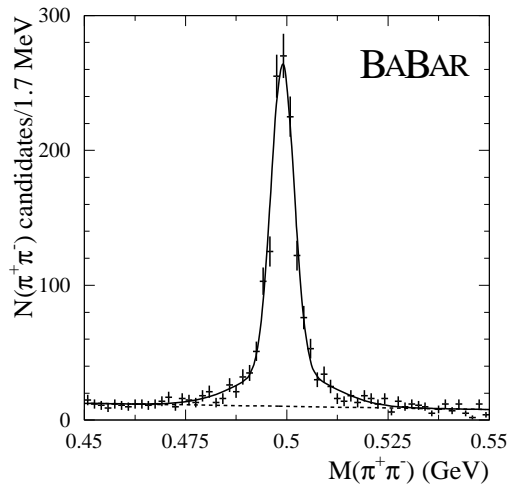}
\caption{$K_S^0$ mass resolution (tight selection). The fit is performed
 with a double Gaussian, the central Gaussian contains 70\%
of the events and has a width of $2.82\pm 0.19$\mevcc.}
\label{pid:neutralhad:ksmass}
\end{center}
\end{figure}


\item
\KS\ $\rightarrow$ \ppz
\par
Non-overlapping \piz\ candidates are combined to construct 
$\KS\rightarrow \piz\piz$ candidates. 
For each candidate with an energy above 800\mev\ 
and a mass between 300 and 700\mevcc, we 
determine the most probable decay vertex along the path 
defined by its momentum vector and the primary vertex. 
The point where the product of the probabilities from the two \piz\
mass constraint fits is a maximum is chosen as the \KS\ decay vertex.
We demand that this point lie in a region between $-10$ and $+40$\cm\ from the 
primary vertex and that the \KS\ mass at that point is in the range
470--536\mevcc. 
This method~\cite{CPLEAR00} 
improves significantly the mass scale and resolution.                                        
A vertex resolution of 5\cm\ along the flight direction is predicted by the simulation. 
The inclusive \KS\ mass peak from real data, exhibiting a resolution 
better than 10\mevcc, is shown in Figure~\ref{fig:kspi0pi0}.

\begin{figure}[!htb]
\begin{center}
\mbox{\epsfig{file=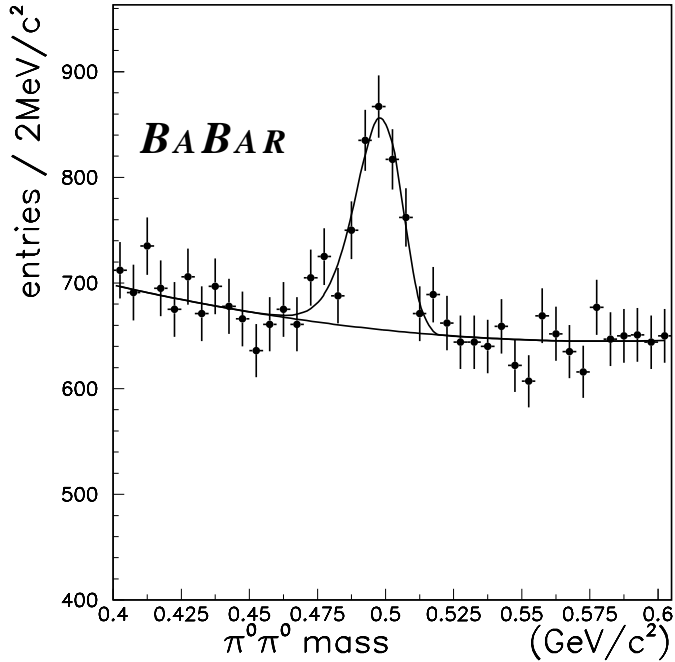,width=0.6\textwidth}}
\end{center}
\caption{\piz\piz\ mass spectrum and \KS\ signal from hadronic events}
\label{fig:kspi0pi0}
\end{figure}

\item
\KL

The EMC and IFR detectors
are both used for the detection of $K_L$
mesons, which are reconstructed as neutral clusters that cannot be
associated with any charged track in the event.
In the EMC, the background consists primarily of photons and high energy
$\pi^0$s. We are able to distinguish $K_L$ clusters from these
backgrounds by using a shower shape analysis, performed by expanding the
shape into a series of Zernike moments. For clusters with an energy
greater than 500\mev, a single Zernike moment (2,0) is sufficient to
discriminate $K_L$ mesons and photons. Between 200 and 500\mev, a
two-dimensional cut ((2,0) and (4,2)) is used. Below 200\mev, we are
unable to sufficiently discriminate between photons and $K_L$ mesons. 
IFR clusters are defined and selected as hits in two or more 
resistive plate chamber layers.
The background is dominantly charged particles and
detector noise. Some splitoffs from charged hadronic showers are missed
by the tracking association, due to the irregular structure of these showers.
We suppress these clusters by rejecting $K_L$ candidate clusters 
close to any track in the event. 
\par
The performance of the $K_L$ identification is demonstrated in Figure
\ref{pid_kl_perf}a, using events from $e^+ e^- \rightarrow \gamma (\phi \rightarrow K_S
K_L)$.  In these events, the position and energy of the $K_L$ can be
constrained from the measured photon and $K_S$ kinematics. Figures
\ref{pid_kl_perf}b and \ref{pid_kl_perf}c
illustrate the performance of the $K_L$ selection.
We observe good agreement between Monte Carlo simulations and
the data. The efficiency for $K_L$ reconstruction agrees
to within the current statistical error, which is approximately $10\%$
for both IFR and EMC selections.

\begin{figure}[hbt]
\epsfig{figure=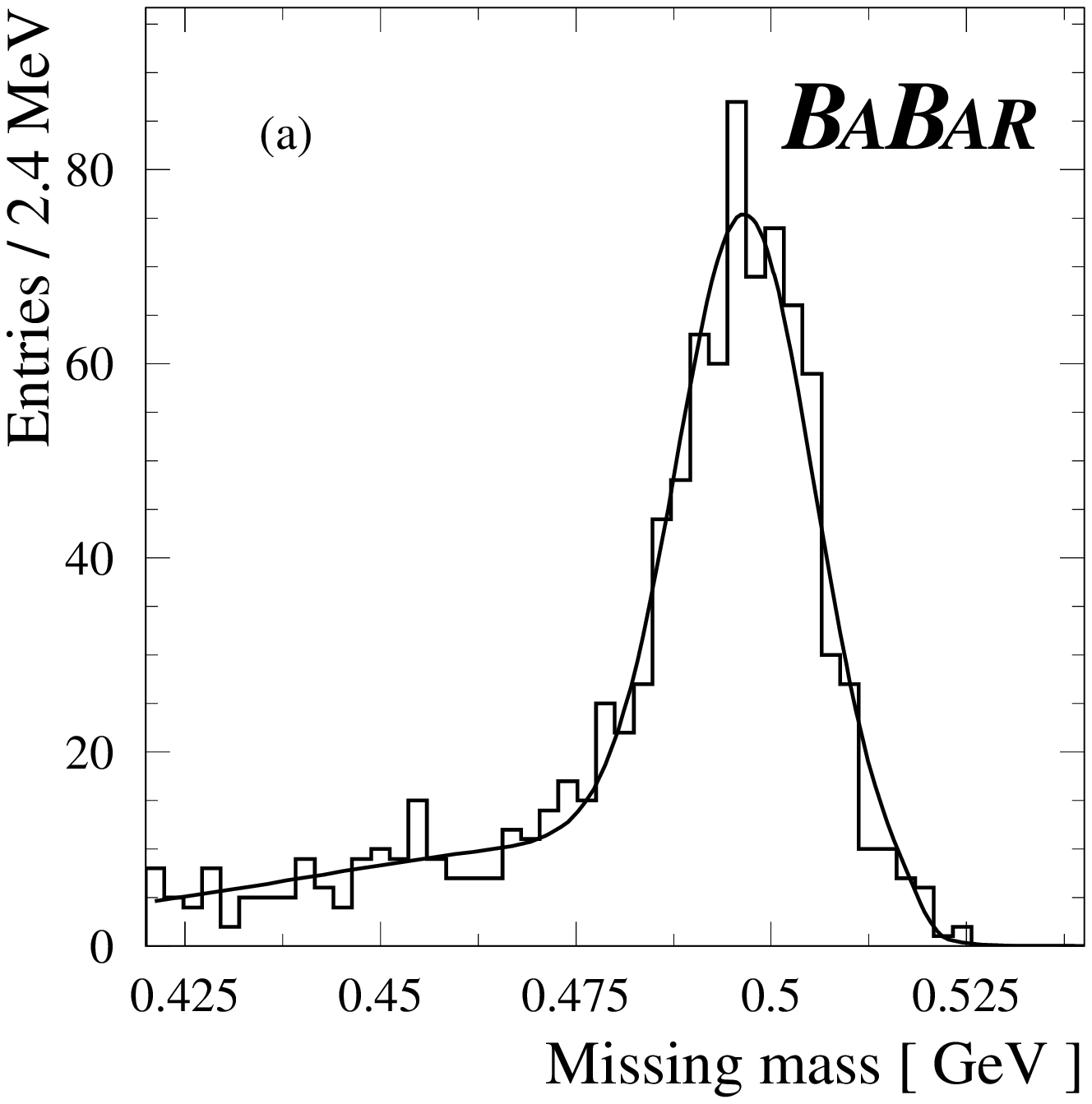,width=0.32\linewidth}
\epsfig{figure=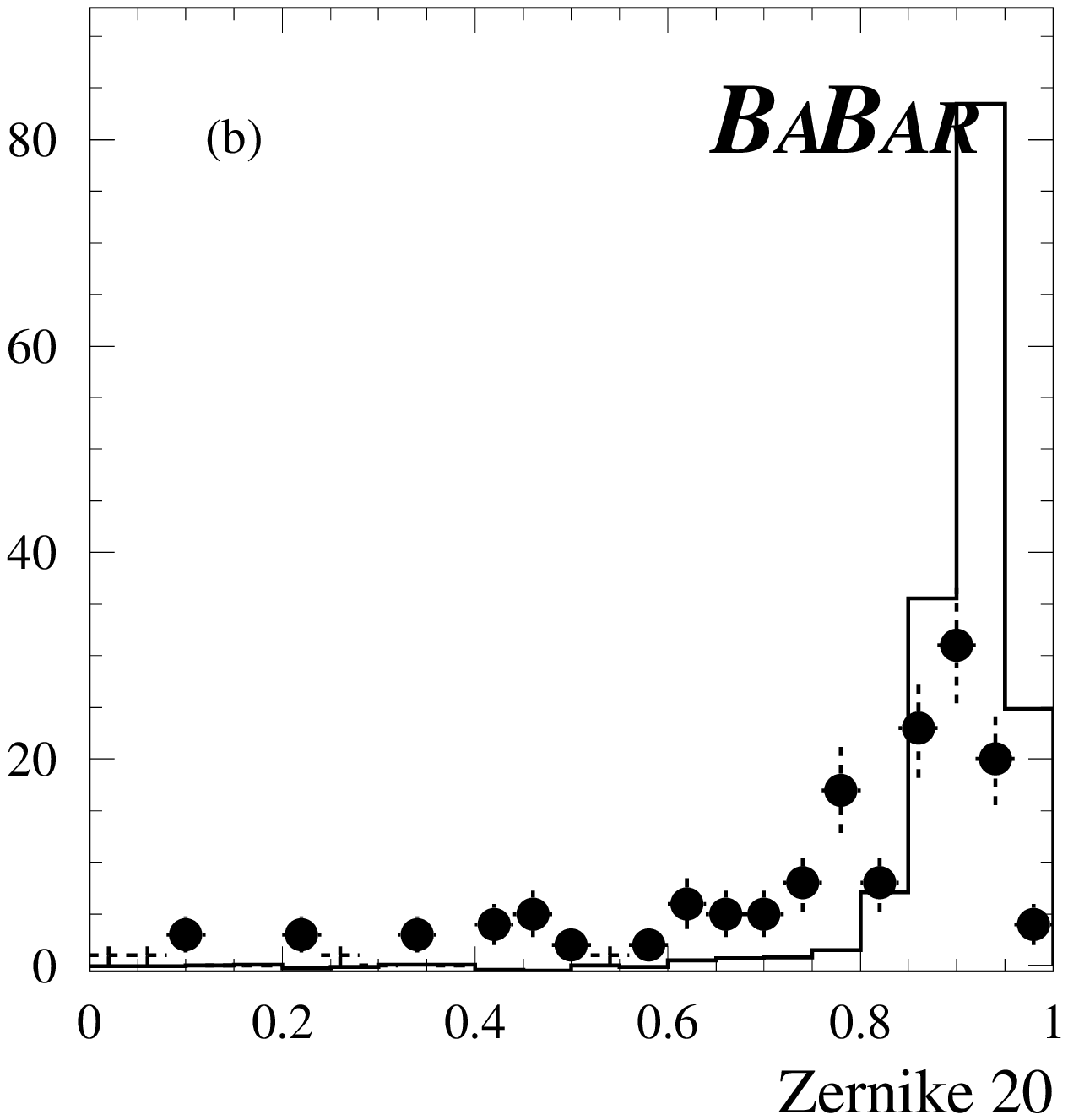,width=0.32\linewidth}
\epsfig{figure=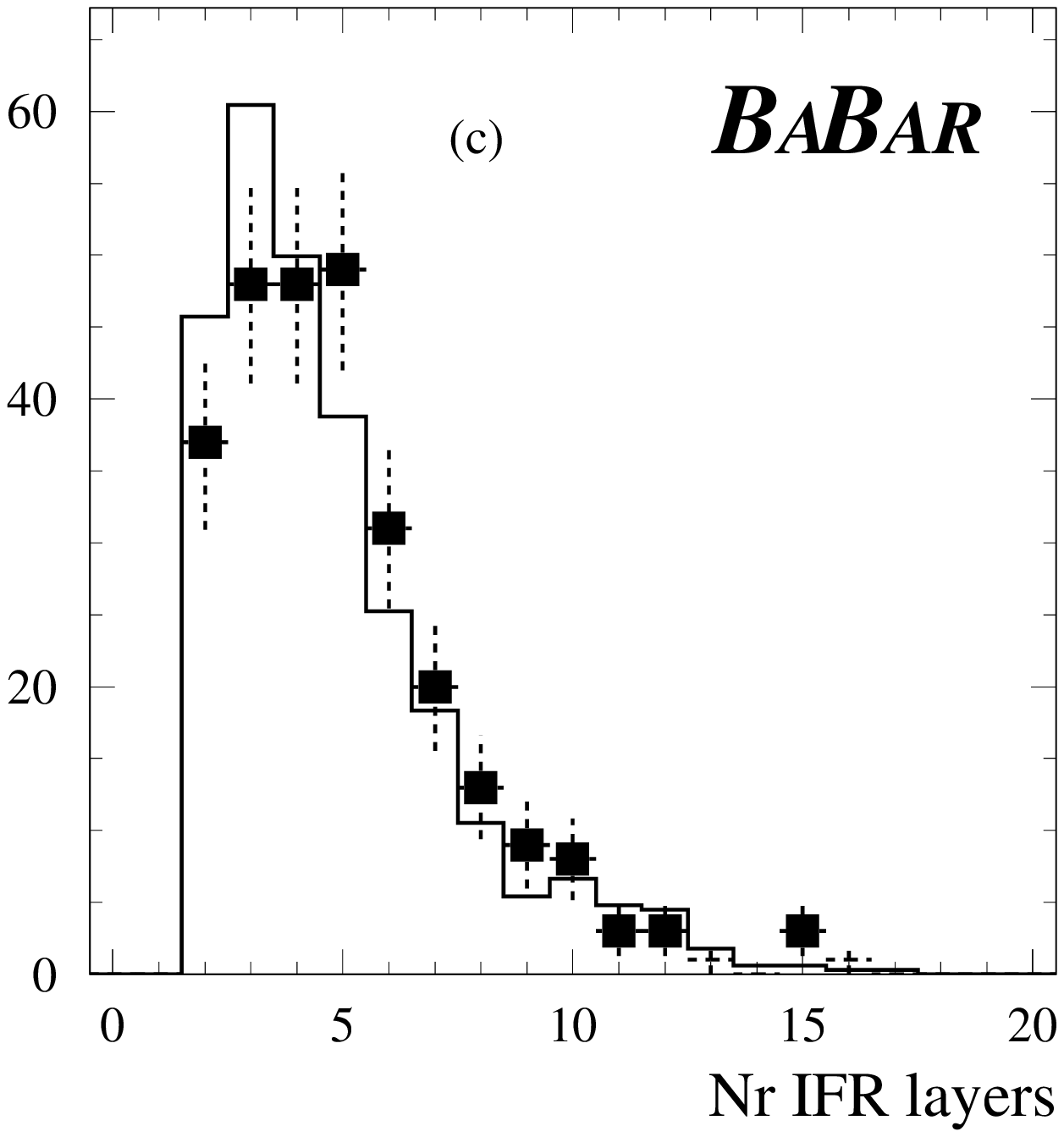,width=0.32\linewidth}
\caption{The missing mass distribution ($p^2_{K_L}$) 
of candidate $\phi \gamma$ events is shown in ({\em a}).
We find $594 \pm 40$ 
$e^+ e^- \rightarrow \gamma (\phi \rightarrow K_S K_L)$ events.
Using these events, ({\em b}) shows the comparison of the Zernike moment (2,0) 
distribution for clusters from $K_L$ mesons (points) and photons (histogram).
A comparison between data (points) and Monte Carlo (histogram) is shown in
({\em c}) for the number of hit layers in the Instrumented Flux Detector (plot ({\em c})
is absolutely normalized).}
\label{pid_kl_perf}
\end{figure}




\end{itemize}


\renewcommand{\secname}{Analysis}          
\section{Common Analysis Issues}
\label{sec:\secname}
\subsection{Centre of Mass Energy and Boost}
\label{sec:\secname:Boost}

The beam energies provided by \pep2\ are obtained with an 
algorithm capable of tracking variations to 
within about $1\mev$, and are used to determine the 
time-dependence of the centre of mass energy.
The absolute energy scale is determined from a sample 
of fully-reconstructed \B\ candidates 
(a recent measurement of the \B\ mass by CLEO~\cite{cleo-mass}
is used).


We define two main coordinate systems:
\begin{itemize}
\item
A \babar\ coordinate system (LAB), linked to the \babar\ detector.  
The $z$ axis is along the DCH axial wires, 
the $y$ axis is vertical (upward) and the $x$ axis points
towards the exterior of the \pep2\ rings. 
The \babar\ $z$ axis and the \pep2\ electron beam 
are not exactly aligned.
The former is tilted in the horizontal plane 
by $19\,{\rm mrad}$ with respect to the beam axis.
\item
A centre-of-mass coordinate system (CMS), a frame where the two
beams have equal energy $E^{*}_b = \sqrt{{\rm s}}/2$ (where
$E^{*}_b$ is the centre-of-mass beam energy).  The $z$ axis
of the CMS lies along the electron beam direction (this is the
relevant choice for most physics quantities).  The Lorentz
transformation from LAB into CMS quantities is
the product of a rotation that aligns the $z^\prime$ axis of the
rotated frame with the boost direction, and a Lorentz boost along the
new $z^\prime$ into the CMS frame. The rotation angles are determined
on a run-by-run basis from the opening angle and flight direction of
$\mu^+\mu^-$ and $e^+e^-$ events, and are verified against the
orientation of the collision spot for the beams. The magnitude of the boost is
obtained from the \pep2\ energies.
\end{itemize}

\subsection{Continuum Rejection Variables}
\label{sec:\secname:Continuum}

We use common shape variables, computed with charged tracks
in the centre of mass frame, to preferenctially reject continuum events.  
The most important of these variables are 
thrust, sphericity and $R2$, the ratio of $2^{\rm nd}$ 
to $0^{\rm th}$ Fox-Wolfram moments.
\begin{itemize}
\item {\bf Thrust:}  The thrust axis, $\hat{T}$, of an event is defined to 
be the  direction that maximizes the sum of the longitudinal 
momenta of the particles.   The value of the thrust $T$ 
must lie in the range 0.5 to 1,
where $T\sim 1$ corresponds to a highly directional event and $T\sim 0.5$
to an isotropic event.
\item {\bf Sphericity:} Sphericity is a measure of the sum of 
squares of transverse momenta for each track with respect to the event axis.
Highly directional events have low sphericity whereas isotropic events tend 
to have sphericities close to one.
\item {\bf Fox-Wolfram moments:}  
The $\ell^{\rm th}$ Fox-Wolfram moment is momentum-weighted sum 
of Legendre polyomial of the $\ell^{\rm th}$ order computed from the cosine 
of the angle between all pairs of tracks.
The ratio $R2$ of Fox-Wolfram $2^{\rm nd}$ 
to $0^{\rm th}$ moments is the variable which provides the best separation 
between \BB\ signal and continuum. Jet-like continuum events tend to 
have higher values of $R2$ than the more spherical \BB\ events. 
\end{itemize}

\subsection{Event Selection and $B$ Counting}
\label{sec:\secname:MultiHadron}

\subsubsection{Fiducial regions}

The acceptance region in polar angle for charged tracks, 
corresponding to full SVT coverage, ranges from
a minimum of 410\mrad\ with respect to the $z$ axis   
in the forward direction to 
602\mrad\ in the backward direction ($0.41<\theta<2.54$).
For neutral clusters, the fiducial region is reduced to an 
angle 732\mrad\ with respect to the $z$ axis in the backward direction
($0.41<\theta<2.41$) to account for the more limited EMC 
coverage in the backward region.

\subsubsection{Multi-Hadron selection}
 
Multi-hadron event selection is designed to have a very high efficiency
for \BB\ events while keeping the systematic uncertainties on the
determination of that efficiency as low as possible.   
The contribution of continuum and $\tau$ pair events 
in the sample is reduced by a cut on $R2$.
Using only tracks and clusters in 
fiducial volume described above, the selection criteria are as follows:
\begin{itemize}
\item
Four or more charged tracks, at least three of which must have associated
DCH information
\item 
Primary vertex within $5\mm$ in $x$ and $y$ of the nominal beam spot position,
\item
$R2$ less than 0.7,
\item 
Total energy greater than 5\gev.
\end{itemize}
The efficiency for \BB\ events is 96\%, while the contamination due
to beam gas, two-photon and tau pair events is approximately 2\%.  

\subsubsection{Muon pair selection}

In the centre of mass frame, 
events containing 
two charged tracks with momenta larger than 2\gevc\ and
4\gevc, respectively, and with angle with respect to the 
beam axis greater than 725\mrad\ are selected. The two tracks
must have an acolinearity smaller than $10^\circ$.  
The mass of the 
pair must be larger than 7.5\gevcc.  The total energy deposited in the 
calorimeter by the two particles must be smaller than 1\gev.  The event is 
rejected if neither track has an associated energy measurement.

\subsection{Offline Luminosity Determination}
\label{sec:\secname:Lumi}

The integrated luminosity recorded by \babar\ is determined offline 
using Bhabha, muon pair, and gamma-gamma events.   
The measurement precision is limited by systematic errors in all cases to 
the several percent level, although this clearly can be improved to 0.5\% or better. 
The systematic uncertainty for the Bhabha measurement is dominated by the efficiency
corrections for the Bhabha veto by the Level 3 trigger, 
and by theoretical 
uncertainties on the differential Bhabha cross section. Likewise,
we are still in the process of comparing and understanding Monte Carlo generators for muon pair
events, including proper treatment of radiative events. 
A better understanding of the impact of occasional hot towers in the EMC will improve
the systematic error on the gamma-gamma luminosity. 
The luminosity determinations are consistent at the several percent level, but are much better than
this for tracking relative luminosity and the variation with beam conditions. 

\subsection{\B\ Counting}
\label{sec:\secname:BCounting}

We estimate the number of \BB\ pairs $N_{\B \Bbar}$ 
from the total number of events passing the 
hadronic event selection $N_{MH}$ and the total number of events passing 
the muon pair selection $N_{\mu \mu}$:
\begin{equation}
N_{\B \Bbar} = N^{{\tt on}}_{MH} - \kappa \times  N^{{\tt off}}_{MH} \times \frac{ N^{{\tt on}}_{\mu \mu} }{  N^{{\tt off}}_{\mu \mu} }  \ \ \ ,
\end{equation} 
where the superscript {\em on} refers to on-resonance data and {\em off} to
off-resonance.  $\kappa$ is a factor close to 1 that corrects for changes in
efficiency and cross section with center of mass energy.  It is estimated by
Monte Carlo to be 0.9962.  The systematic error on $N_{\BB}$ is 1.7\%,
dominated by the variation between running periods of the ratio 
$N^{off}_{MH}/N^{off}_{\mumu}$.

Branching fraction
measurements frequently use the number of {\em produced} \BB\ pairs.  The
efficiency for \BB\ events to pass the selection criteria is determined by
Monte Carlo to be 96.0\%.  The total systematic error on the number of produced
\BB\ pairs is 3.6\%.

\subsection{Exclusive \B\ Reconstruction Variables}
\label{sec:\secname:BVariables}
\subsubsection{Kinematic variables}

A number of kinematic quantities could be used to 
characterize the reaction $\epem\to\B\Bbar$ where one of the 
$B$ mesons is fully reconstructed. However, our goal is to select
a pair of such kinematic variables, having little
correlation between them, that make
use of the maximum available information for optimal background 
rejection.

At the \FourS\ resonance the \B\ mesons are produced with 
very small $Q$ values. As a result their
center-of-mass energy $E^{*}$ and momentum $p^{*}$  
are very sensitive to fluctuations of the center-of-mass beam energy 
$E^{*}_b = \sqrt{{\rm s}}/2$. 
In contrast, 
the variable ${\rm \Delta}E$ defined as:
\begin{equation}
	{\rm \Delta}E = E^{*} - E^{*}_b \ \ ,
\end{equation}
is relatively insensitive to $E^{*}_b$ fluctuations.  

The distribution of ${\rm \Delta}E$ is peaked at zero for  $\epem\to\B\Bbar$ 
events, 
and its width is 
governed in most cases  by the beam energy measurement error $\sigma^2_E$.
It is not necessary to 
boost the \B\ candidate to the center-of-mass frame to compute 
${\rm \Delta}E$, as can be seen by writing    
${\rm \Delta}E$ in an explicitely 
Lorentz invariant form:
\begin{equation}
	{\rm \Delta}E = \frac{  2 \,  \widetilde{P} . \widetilde{P}_i - {\rm s} }{ 2 \sqrt{ {\rm s } } } \ \ \ ,
\end{equation}
where $\widetilde{P} $ 
and $\widetilde{P}_i $ are the Lorentz vectors representing the 
\B\ candidate four momentum and the initial-state four momentum
respectively.

The ${\rm \Delta} E$ variable is used in conjunction with either one of
 the following two mass variables:
\begin{itemize}
\item
{\bf Beam-energy substituted mass} $\mes$, defined 
as:
\begin{equation}
\label{eq:mse}
      \mes =   \sqrt{  \left(  \frac{ \frac{1}{2} {\rm s} + \overrightarrow{ p} . \overrightarrow{ p}_{\! i}}{E_i} \right)^2  -  p^2 } \ \ \ ,
\end{equation}
where $E_i$ and $\overrightarrow{ p}_{\! i}$  are  the total energy
and the three momentum of the initial state in the laboratory frame, 
and $\overrightarrow{ p}$ is  the three momentum of the  \B\ candidate 
in the same frame.
At a symmetric machine 
(where $\overrightarrow{ p}_{\! i} = \overrightarrow{0}$) 
this variable is obtained simply by substituting the beam energy $ E^{*}_b$ from the measured  \B\ candidate energy $ E^{*}$:
\begin{equation}
	\mes = \sqrt{  E^{* 2}_b - p^{* 2} }  \ \ \ .
\end{equation}
This is the definition of the variable used 
in ARGUS and CLEO publications under the name 
``beam-constrained'' mass.  
The advantage of using definition of Eq.~\ref{eq:mse} at an
asymmetric machine, where one needs to assign masses to candidates 
in order to boost to the center of mass frame,
is that $\mes$ is 
computed in the laboratory frame without 
any prior commitment to the identification of particles 
among the \B\ daughters.  
\item
{\bf Beam-energy constrained mass} \mec, defined as:
\begin{equation}
      \mec = \sqrt{  \widehat{E}^{\, 2} - \widehat{p}^{\, 2} } \ \ \ ,
\end{equation}
where $\widehat{E}$ and $\widehat{p}$ are center-of-mass frame quantities 
obtained by performing a kinematic fit with the $E^* = E_b$ constraint. 
\end{itemize}

The two choices 
$({\rm \Delta} E, \, \mes )$ and $({\rm \Delta} E, \, \mec )$ 
are almost equivalent.  
For certain analyses, it may be helpful to 
exploit the property that $ \mes$ does not require daughter mass assignments,
while  the  kinematical
fit performed for $\mec$ 
makes optimal use of the information collected by the detector.  
in addition, the pair $({\rm \Delta} E, \, \mec )$ 
has the smallest correlation.

\subsubsection{Background parametrisation}

To describe the background shape in beam-energy constrained mass plots, 
we use the ARGUS distribution~\cite{Argus} given by:
\begin{equation}
 {\cal A}( \, m \, | \, m_0,\,  c \,  ) = \frac{{\rm \theta}( m < m_0 )}{N} \times m \sqrt{ 1 - ( m / m_0 )^2 } 
 \times \exp{ \left[ \, c \, ( 1 - m/m_0 )^2 \,  \right] } \ \ \ ,
\end{equation}
where $m_0$ represents the kinematic upper limit and is generally 
held fixed at the center-of-mass beam energy  $ E^{*}_b$.

 
\renewcommand{\secname}{Conclusions}          
\section{Conclusions and Prospects}
Since the start of running at \pep2\ in May 1999, the \babar\ experiment has
accumulated over 12.7\invfb\ of data. Many of the detector sub-systems have
reached their design goals and significant progress has been made towards
understanding the performance of the detector. Several preliminary
measurements with the \babar\ detector are to be presented at this
conference. We anticipate having a data set equivalent to a significantly larger integrated
luminosity by the end of this year. Thus, the results presented here are just 
the beginning of an exciting program of physics with \babar\ at
\pep2.
In future years, we expect to be able to undertake detailed studies of \CP\ violation 
in the \B\ meson system. 

\label{sec:\secname}


\section{Acknowledgments}
\label{sec:Acknowledgments}
We are grateful for the contributions of our \pep2\ colleagues in
achieving the excellent luminosity and machine conditions
that have made this work possible.
We acknowledge support from the
Natural Sciences and Engineering Research Council (Canada),
Institute of High Energy Physics (China),
Commissariat \`a l'Energie Atomique and
Institut National de Physique Nucl\'eaire et de Physique des Particules
(France),
Bundesministerium f\"ur Bildung und Forschung
(Germany),
Istituto Nazionale di Fisica Nucleare (Italy),
The Research Council of Norway,
Ministry of Science and Technology of the Russian Federation,
Particle Physics and Astronomy Research Council (United Kingdom), the
Department of Energy (US),
and the National Science Foundation (US). In addition, individual support 
has been received from the Swiss 
National Foundation, the A. P. Sloan Foundation, the Research Corporation,
and the Alexander von Humboldt Foundation.
The visiting groups wish to thank 
SLAC for the support and kind hospitality
extended to them.

\end{document}